\documentclass[lettersize,journal]{IEEEtran}
\usepackage{amsmath,amsfonts}
\usepackage{algorithm}
\usepackage{algcompatible}
\usepackage{array}
\usepackage[caption=false,font=normalsize,labelfont=sf,textfont=sf]{subfig}
\usepackage{textcomp}
\usepackage{stfloats}
\usepackage{url}
\usepackage{verbatim}
\usepackage{graphicx}
\usepackage{cite}
\usepackage{caption}
\usepackage{soul}
\usepackage{comment}
\usepackage{xcolor}
\usepackage{booktabs}
\newif\ifshow
\showfalse

\newif\ifblue
\bluetrue
\newif\ifred
\redtrue
\newif\ifminor
\minortrue
\begin{document}
% \begingroup
% % \let\clearpage\relax 
% \onecolumn 
% \input{response}
% \endgroup

\renewcommand\thefigure{\arabic{figure}}
\setcounter{figure}{0}

\twocolumn
\title{Progressive Frame Patching for FoV-based Point Cloud  Video Streaming}

\author{Tongyu~Zong, Yixiang~Mao, Chen~Li, 
        Yong Liu,~\IEEEmembership{Fellow,~IEEE,}
        and~Yao~Wang,~\IEEEmembership{Fellow,~IEEE}
        % <-this % stops a space
\thanks{T. Zong, Y. Mao, C. Li, Y. Liu and Y. Wang are with the Department
of Electrical and Computer Engineering, New York University, Brooklyn,
NY, 11201 USA e-mail: (\{tz1178,yixiang.mao,chen.lee,yongliu,yw523\}@nyu.edu).}% <-this % stops a space
% \thanks{This paper was produced by the IEEE Publication Technology Group. They are in Piscataway, NJ.}% <-this % stops a space
% \thanks{Manuscript received April 19, 2021; revised August 16, 2021.}}
}

% The paper headers
% \markboth{Journal of \LaTeX\ Class Files,~Vol.~14, No.~8, August~2021}%
% {Shell \MakeLowercase{\textit{et al.}}: A Sample Article Using IEEEtran.cls for IEEE Journals}

% \IEEEpubid{0000--0000/00\$00.00~\copyright~2021 IEEE}
% Remember, if you use this you must call \IEEEpubidadjcol in the second
% column for its text to clear the IEEEpubid mark.

\maketitle

\begin{abstract}
  Many XR applications require the delivery of volumetric video to users with six degrees of freedom (6-DoF) movements. Point Cloud has become a popular volumetric video format. A dense point cloud consumes much higher bandwidth than a 2D/360 degree video frame. User Field of View (FoV) is more dynamic with 6-DoF movement than 3-DoF movement. To save bandwidth, FoV-adaptive streaming predicts a user's FoV and only downloads point cloud data falling in the predicted FoV. However, it is vulnerable to FoV prediction errors, which can be significant when a long buffer is utilized for smoothed streaming. In this work, we propose a multi-round progressive refinement framework for point cloud video streaming. Instead of sequentially downloading point cloud frames, our solution simultaneously downloads/patches multiple frames falling into a sliding time-window, leveraging the inherent scalability of octree-based point-cloud coding. The optimal rate allocation among all tiles of active frames are solved analytically using the heterogeneous tile rate-quality functions calibrated by the predicted user  FoV. Multi-frame downloading/patching simultaneously takes advantage of the streaming smoothness resulting from long buffer and the FoV prediction accuracy at short buffer length. We evaluate our streaming solution using simulations driven by real point cloud videos, real bandwidth traces, and 6-DoF FoV traces of real users. Our solution is robust against the bandwidth/FoV prediction errors, and can deliver high and smooth view quality in the face of bandwidth variations and dynamic user and point cloud movements.
\end{abstract}


\section{Introduction}
\label{sec:intro}
Point cloud video streaming will take telepresence to the next level by delivering full-fledged 3D information of the remote scene and facilitating six-degree-of-freedom (6-DoF) viewpoint selection to create a truly immersive experience. 
%With the  fascinating advances in the key enabling technologies, ranging from high-fidelity volumetric capturing, 3D computer vision processing, high-throughput and low-latency 5G/6G data delivery, to economically-viable volumetric rendering, we are now at the verge of  completing the puzzle of teleporting holograms of real-world humans/creatures/objects through the global Internet to realize the full potentials of  Virtual/Augmented/Mixed Reality (XR in short).
Leaping from planar (2D) to volumetric (3D) video poses significant communication and computation challenges. 
%for volumetric video processing and delivery. 
A point cloud video consists of a sequence of frames that characterize the motions of one or multiple physical/virtual objects. Each frame is a 3D snapshot, in the form of point cloud captured by a 3D scanner, e.g., LiDAR camera, or a camera array using photogrammetry. A high-fidelity point cloud frame of a single object can easily contain one million points, each of which has three 32-bit Cartesian coordinates and three 8-bit color attributes. At 30 frames/second, the raw data rate of a point cloud video (PCV) of a single object reaches 3.6 Gbps. The raw data rate required to describe a 3D scene with multiple objects increases proportionally.  
% Meanwhile, compressing high-volume PCV %\cite{li_jia2021convolutional} 
% for bandwidth-efficient transmission can consume substantial 
% computation resources on the source side. 

Meanwhile, on the receiver side, a PCV viewer enjoys 6-DoF viewpoint selection through translational ({\it x, y, z}) and head rotational ({\it pitch, yaw, roll}) movements. 
% %Viewer movements and PCV object motions are projected to the same virtual/augmented environment to create the illusion of co-existence and interaction.  
Given the relative position between the object and the viewer's viewpoint, the compressed PCV will be decompressed and rendered on a 2D or 3D display
%\footnote{3D displays, such as light-field or laser-based hologram displays, are still not accessible for most users. We will consider FoV-based 2D PCV rendering on flatscreens widely available on computers, TVs, mobile devices, and head-mount-displays. Our designs are applicable for PCV streaming to 3D displays when  available.  }
, which also consumes significant computation resources. %\cite{li_li2022plenoptic}. 
Furthermore, to facilitate seamless interaction and avoid motion sickness, the rendered view has to be delivered with short latency (e.g. $<$20 ms) after the viewer movement, the so called Motion-to-Photon (MTP) latency constraint \cite{cuervo2018creating_motion_to_photon_latency}. As a result,  PCV not only consumes more bandwidth and computation resources, it is additionally subject to stringent end-to-end latency budget for processing and delivery. 

{\it The goal of this work is to address the high-bandwidth, high-complexity and low-latency challenges of point cloud video by designing a novel progressive refinement streaming framework, where multiple frames falling into a sliding time window are patched simultaneously for multiple rounds, leveraging the inherent scalability of octree-based point-cloud coding.} We will focus on video-on-demand applications, leaving the live PCV streaming and the ultimate challenge of realtime two-way interactive PCV streaming for future research. 

Towards developing this progressive streaming framework, we made the following contributions:
\begin{enumerate}
\item We design a novel {\it sliding-window based  progressive streaming framework} to gradually refine the spatial resolution of each tile as its playback time approaches and the FoV prediction accuracy improves. We investigate the trade-off between the need of long-buffer for absorbing bandwidth variations and the need of short-buffer for accurate FoV prediction.
% We investigate the {\it computation-communication trade-off in progressive downloading and post-processing}. 
\item We propose a novel point cloud tile rate-quality model that reflects the Quality of Experience (QoE) of PCV viewers both theoretically and empirically.

\item The tile rate allocation decision is  formulated as a utility maximization problem. To get the optimal solution, an analytical algorithm based on Karush–Kuhn–Tucker (KKT) conditions is developed.
% \item We further formulate the rate allocation problem as an optimal control decision making problem and solve it with iterative Linear Quadratic Regulator (iLQR), which further enhances the Quality of Experience (QoE) performance compared to KKT-based algorithm due to its calculation of future impacts of every action.

\item The superiority of our proposed algorithm is evaluated by QoE model as well as 2D rendered view and PSNR/SSIM, using PCV streaming simulations driven by  real point cloud video codec and real-world bandwidth traces. 
% with {\bf rendered visual results ??? to be updated ???}
\end{enumerate}

\section{Related Work}
\label{sec:related}
%Impressive research have been accomplished for coding~\cite{360Coding1,360Coding2,liu_zhou2022exploring,Winstein_hsiao2022towards},  streaming~\cite{Jiasi2017,Nahrstedt_zink2019scalable,jiang_guan2019pano,qiu_he2018rubiks,360Delivery1,li2019very,Duanmu:SigComm,Liyang_MMSys18,sun2019twotier,mao2020low, sun2020flocking,liu_jiang2020qurate,Jacob2,qian_wang2022salientvr}, FoV prediction~\cite{bao2016shooting,Zhisheng_MM18,Nahrstedt_park2020seaware}, and edge-assisted delivery~\cite{mao_dai2019view,Chakareski_liu2020delivering,Nahrstedt_sarkar2021l3bou} of 360$^o$  video recently. 
%We have applied a subset of the previously stated design principles to  360$^o$  video coding and  streaming~\cite{Duanmu:ISCAS, Duanmu:SigComm,Liyang_MMSys18,sun2019twotier,mao2020low, sun2020flocking,Flocking-TMM22,li2019very}. 
%However, a 360$^o$  video only covers the whole viewing sphere captured from a center position, 
%records 2D projection along each direction from the center, 
%and is typically projected onto a 2D plane  and processed as a planar video. A viewer mostly makes 3-DoF head rotational movements. 
%rotate her head in 2-DoF (pitch and yaw). 
%A PCV records the 3D space directly and allows 6-DoF viewpoint selection from different positions and angles.  

% {\bf ??? include more recent papers on PCV coding and streaming ???? }

PCV coding and streaming is extremely different from traditional 2D video as shown in~\cite{akhshabi2011experimental}. Several PCV coding and streaming systems, e.g.~\cite{chen2023patchvvc,wang2023vqba,zhang2023g,wang2023hermes,park2019rate,qian_han2020vivo,groot2020,qian_zhang2022yuzu,qian_liu2022vues,wang2021qoe,wang2022qoe,rossi2023extending,subramanyam2020user,kammerl2012real,shi2023enabling,mekuria2016design,zhang2021efficient,sheng2021deep,wang2021lossy,mao2022learning,liu2020fuzzy,van2019towards,gao2022fras,li2022optimal,li2023towards,rudolph2023rabbit,zhang2023yoga}, have shown promises.
% the design space of PCV coding, streaming, and FoV prediction is widely open to be holistically explored.
%They precode the PCV into multiple rate versions for rate adaptation. 
For example, \cite{chen2023patchvvc} proposes patchVVC, a point cloud patch-based compression framework that is FoV adaptive, and achieves both high compression ratio and real-time decoding. The authors of YOGA~\cite{wang2023vqba} propose a
frequency-domain-based profiling method to transform point cloud into a vector before estimating the compressed bitrate for geometry data and color information using linear regression. Then they perform rate allocation between geometry map and attribute map in the context of V-PCC~\cite{graziosi2020overview}. The Hermes system~\cite{wang2023hermes} adopts an implicit correlation encoder to reduce bandwidth consumption and proposes a hybrid streaming method that adapts to dynamic viewports. The ViVo system \cite{qian_han2020vivo} adopts frame-wise k-d tree representation using DRACO \cite{Draco}, which is simple but less efficient than the recently established MPEG G-PCC standard \cite{graziosi2020overview}. It uses tile-based coding to enable FoV adaptation and employs non-scalable multiple-rate versions of the same tile for rate adaptation. Moreover, it only performs prefetching over a short interval ahead, leading to limited robustness to network dynamics. Another study \cite{qian_zhang2022yuzu} considered delivering low-resolution PCV and used deep-learning models to enhance the resolution on  the receiver, which enhances coding efficiency but does not facilitate FoV adaptation. \cite{hosseini2018dynamic} extends the concept of dynamic adaptive streaming over HTTP (DASH) towards DASH-PC (DASH-Point Cloud) to achieve bandwidth and view adaptation for PCV streaming. It proposes a clustering-based sub-sampling approach for adaptive rate allocation, while our work is able to optimize the rate allocation explicitly based on the FoV prediction accuracy and tile utility function. \cite{mekuria2016design} presents a generic and real-time time-varying point cloud encoder, which not only adopts progressive intra-frame encoding, but also encodes inter-frame dependencies based on an inter-prediction algorithm. To better match user's FoV, the authors of \cite{li2022optimal} propose a hybrid visual saliency and hierarchical clustering based tiling scheme. They also propose a joint computational and communication resource allocation scheme to optimize the QoE. Most of the existing studies focused on streaming PCV of a single object without \textit{explicit} rate control. In this work, we assume octree-based scalable PCV coding which simultaneously enables spatial scalability and FoV adaptation with fine granularity. Our proposed progressive streaming framework takes full advantage of scalable  PCV coding to enhance streaming robustness. Besides, there are also several study focusing on quality assessment of PCV such as \cite{ak2024basics,ak2024toolkit}.

\cite{park2019rate} is the first work to propose a window-based streaming buffer that supports rate update for all tiles in the buffer and the tile rate allocation problem is solved by a heuristic greedy algorithm. 
% However, it didn't exploit scalable coding for point cloud tiles and the tile rate allocation problem is solved by a heuristic greedy algorithm that may not obtain the optimal solution. 
Furthermore, it proposes a tile utility function to model the true user QoE. However, it is not clear how the coefficients of the utility model are obtained. We  formulate the PCV streaming problem with the well developed concept of scalable octree coding, and propose a more refined utility model that better reflects viewer's visual experience when viewing a point cloud tile from certain distance. To achieve this, we coded several test PCVs using MPEG G-PCC and found that the level of detail (LoD) is logarithmically related to the rate.  Due to tile diversity, the data rate needed to code octree nodes up to a certain level is highly tile-dependent. We fit the coefficients of this logarithmic mapping from rate to LoD and model the tile utility as a function of angular resolution to match the shape of subjective utility-(rate, distance) curve in~\cite{qian_han2020vivo}. We also develop a new tile rate allocation algorithm based on the KKT Condition that outperforms the heuristic algorithm of \cite{park2019rate} in our evaluation.

\section{FoV-adaptive Coding and Streaming Framework}
\label{sec:formulation}

% \begin{figure*}[htbp]
% \centering 
%     \begin{subfigure}[b]{width=.3\linewidth}
%         % \centering
%         \includegraphics[width=0.3\linewidth,height=1.6in]{figs/cube.eps}
%         \caption{Point Cloud in Voxel Space}
%         \label{fig:voxel}
%     \end{subfigure}
%     % \hfill
%     \begin{subfigure}[b]{width=.3\linewidth}
%         % \centering
%         \includegraphics[width=0.3\linewidth,height=1.6in]{figs/Octree.eps}
%         \caption{Octree Representation}
%         \label{fig:octree1}
%     \end{subfigure}
%     % \hfill
%     \begin{subfigure}[b]{width=.36\linewidth}
%         % \centering
%         \includegraphics[width=0.36\linewidth,height=1.6in]{figs/Tile_Tree.eps}
%         \caption{Tile-based Scalable PCV Frame Coding}
%         \label{fig:tiletree}
%     \end{subfigure}

% \caption{Octree-based Scalable and FoV-adaptive Point Cloud Coding}
% \label{fig:octree}
% \end{figure*}

\begin{figure*}[!t]
    \centering 
    \subfloat[\footnotesize Point Cloud in Voxel Space]{\includegraphics[width=.3\linewidth,height=1.6in]{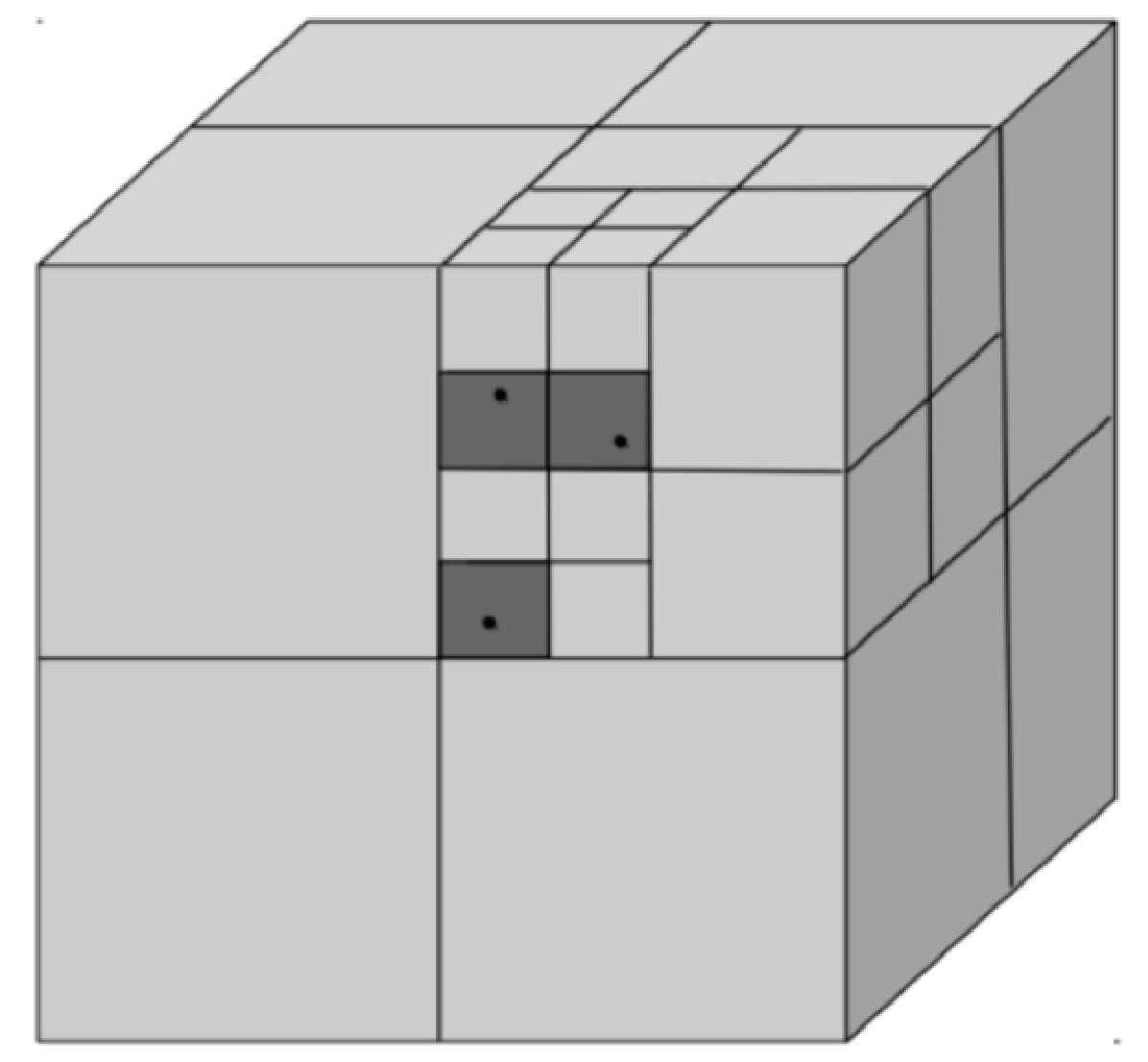}%
    \label{fig:voxel}}
    \hfil
    \subfloat[\footnotesize Octree Representation]{\includegraphics[width=.3\linewidth,height=1.6in]{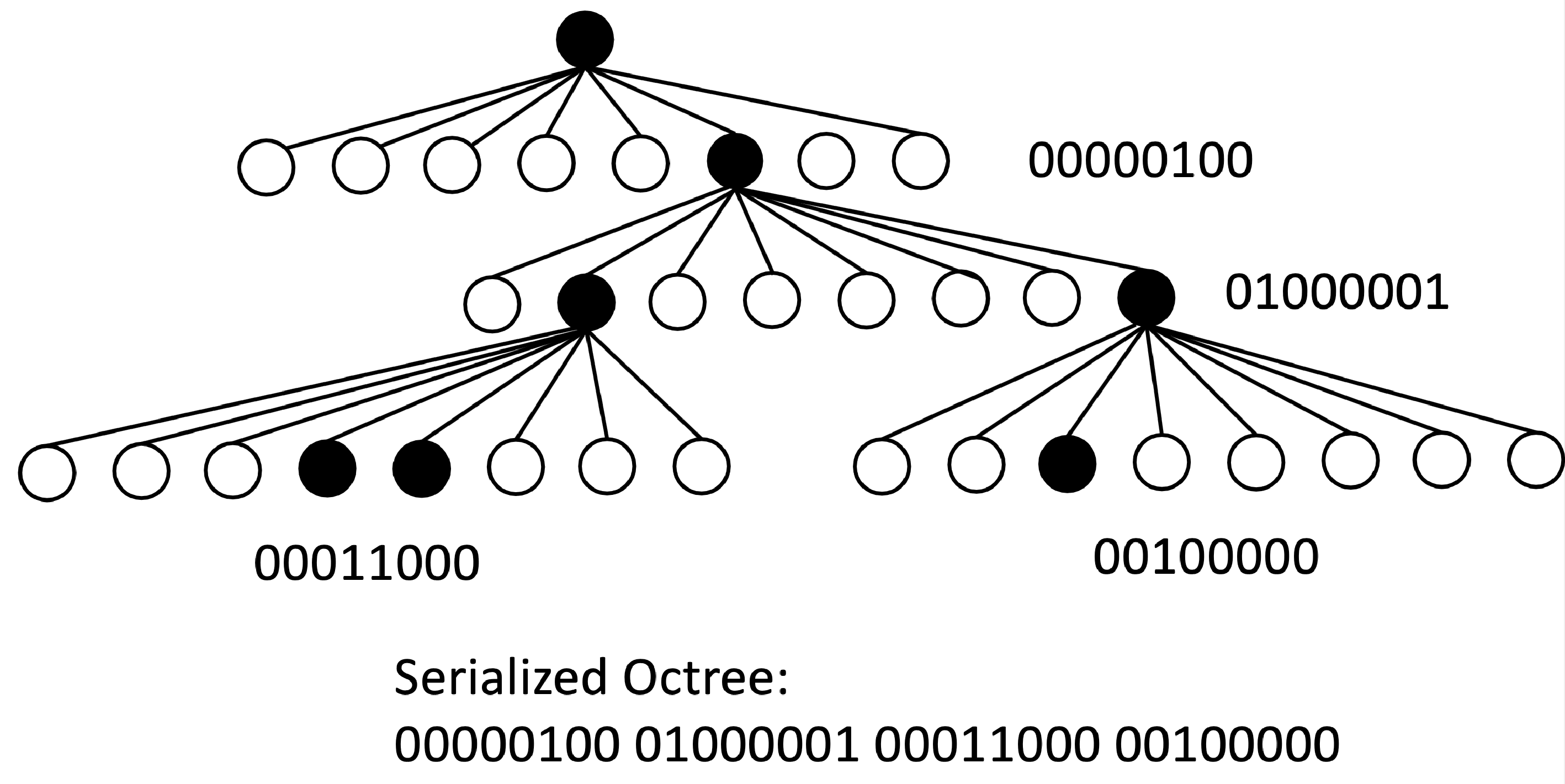}%
    \label{fig:octree1}}
    \hfil
    \subfloat[\footnotesize Tile-based Scalable PCV Frame Coding]{\includegraphics[width=.36\linewidth,height=1.6in]{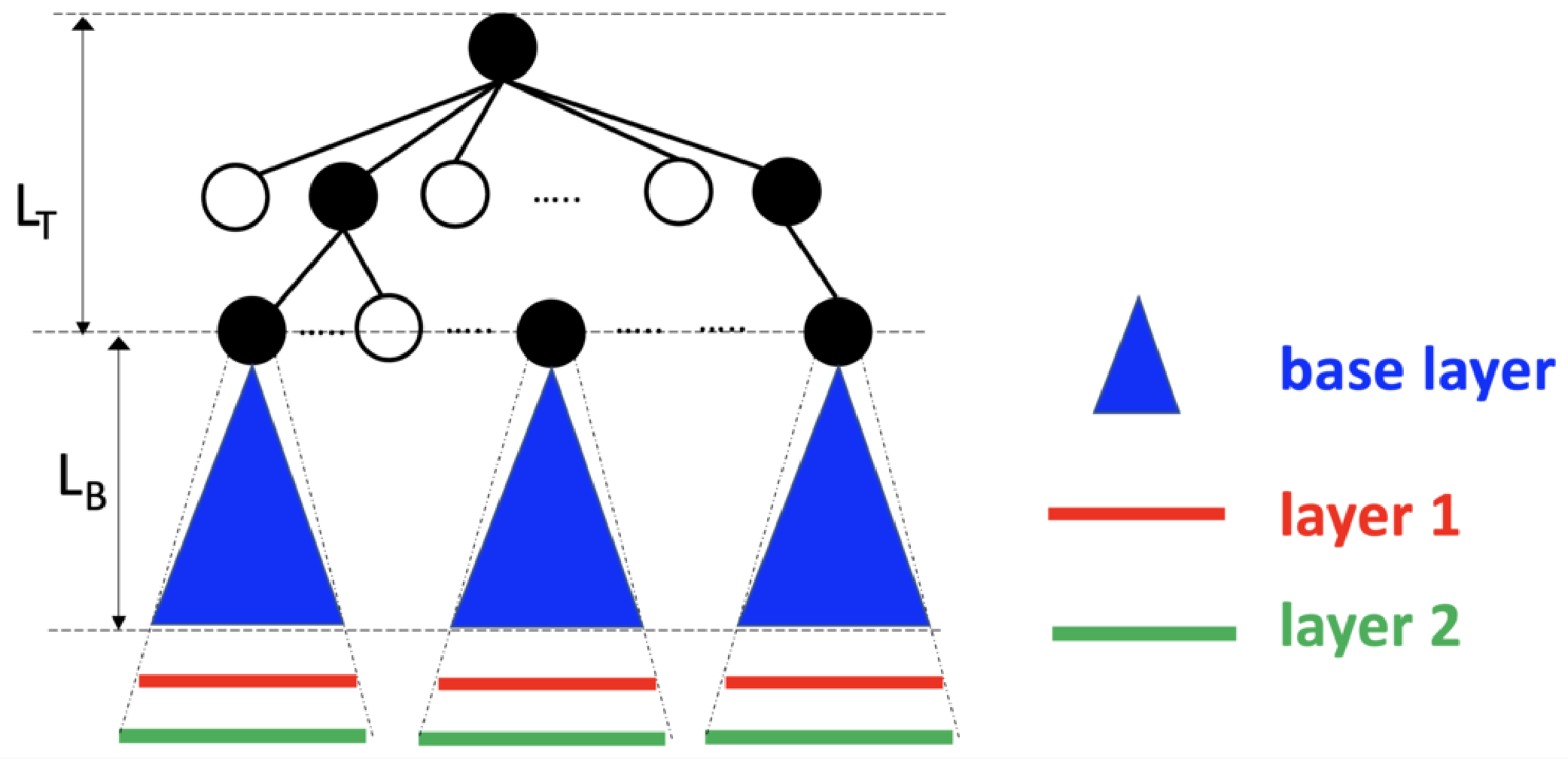}%
    \label{fig:tiletree}}
\caption{Octree-based Scalable and FoV-adaptive Point Cloud Coding}
\label{fig:octree}
\end{figure*}

\subsection{Octree-based Scalable PCV Coding}
MPEG has considered point cloud coding (PCC) and recommended two different approaches \cite{graziosi2020overview}. The V-PCC approach projects a point cloud frame into multiple planes and assembles the projected planes into a single 2D frame, and code the resulting sequence of frames using previously established video coding standard HEVC. The G-PCC approach represents the geometry of all the points using an octree, and losslessly represents the octree using context-based entropy coding. The color attributes are coded following the octree structure as well using a wavelet-like transform. By leveraging the efficiency of HEVC, especially in exploiting temporal redundancy through motion compensated prediction, the V-PCC approach is more mature and is currently more efficient than the G-PCC for dense point clouds. However, V-PCC does not allow selective transmission of regions intersecting with the viewer's FoV or objects of interests. It is also not efficient for sparse clouds such as those captured by LiDAR sensors.  The spatial scalability can be indirectly enabled by coding the projected 2D video using spatial scalability, but this does not directly control the distance between the points and cannot be done at the region level.

A point cloud can be coded using an octree which is recursively constructed: the root node represents the entire space spanned by all the points, which is equally partitioned along the three dimensions into $2\times 2\times 2$ cubes; the root has eight children, each of which is the root of a sub-octree representing each of the eight cubes. The recursion stops at either an empty cube, or the maximum octree level $L$.  Each sub-octree rooted at level $l$ represents a cube with side lengths that are $1/2^l$ of the side lengths of the original space. To code the geometry, the root node records the $(x,y,z)$ coordinates of the center of the space, based on which the coordinates of the center of any cube at any level are uniquely determined. Consequently, each non-root node only needs one bit to indicate whether the corresponding cube is empty or not. For the octree in Fig.~\ref{fig:octree1}, the nodes with value $1$ at the three levels represent one, two and three non-empty cubes with side lengths of $1/2$, $1/4$ and $1/8$ in Fig.~\ref{fig:voxel}, respectively. All the non-empty nodes at level $l$ collectively represent the geometry information of the point cloud at the spatial granularity of $1/2^l$. This makes octree {\it spatially scalable}: with one more level of nodes delivered, the spatial resolution doubles (i.e., the distance between points is halved). The color attributes of a point cloud are coded following the octree structure as well, with each node stores the average color attributes of all the points in the corresponding cube. Scalable color coding can be achieved by only coding  the difference between a node and its parent. The serialized octree can be further losslessly compressed using entropy coding. With MPEG G-PCC, at each tree level $l$, the  status of a node ($0$ or $1$) is coded using context-based arithmetic coding, which uses a context consisting of its neighboring siblings that have been coded and its neighboring parent nodes to estimate the probability $p$ that the node is $1$.

% {\bf Yixiang: ...add description about the PCV compression scheme used in the evaluation.... entropy coding}

\subsection{Tile-based Coding and Streaming}
 Considering the limited viewer Field-of-View (FoV) (dependent on the 6-DoF viewpoint), occlusions between objects and parts of the same object, and the reduced human visual sensitivity at long distances, only a subset of points of a PCV frame are {\it visible and discernible} to the viewer at any given time. FoV-adaptive PCV  streaming significantly reduces PCV bandwidth consumption by only streaming points visible to the viewer at the spatial granularity that is discernible at the current viewing distance. To support FoV-adaptive streaming, octree nodes are partitioned into slices that are selectively transmitted for {\bf FoV adaptability} and {\bf spatial scalability}. Each slice consists of a subset of nodes at one tree level that are independently decodable provided that the slice containing their parents is received. Without considering FoV adaptability, one can simply put all nodes in each tree level into a slice to achieve spatial scalability. The  sender will send slices from the top level to the lowest level allowed by the rate budget. To enable FoV adaptability, a popular approach known as tile-based coding is to partition the entire space into non-overlapping 3D tiles and code each tile independently using a separate tile octree. Only tiles  falling into the predicted FoV will be transmitted. To enable spatial scalability within a tile, we need to put the nodes at each level of the tile octree into a separate slice. As illustrated in Fig.~\ref{fig:tiletree}, each sub-octree rooted at level $L_T$ represents a 3D-tile with side length of $1/2^{L_T}$. Within each tile sub-octree, nodes down to the $L_B$ level are packaged into a base layer slice, and nodes at the lower layers are packaged into additional enhancement layer slices. 

When the streaming server is close to the viewer, one can conduct reactive FoV-adaptive streaming: the client collects and uploads the viewer's FoV to the server; the server renders the view within the FoV, and streams the rendered view as a 2D video to the viewer. To facilitate seamless interaction and avoid motion sickness, the rendered view has to be delivered with short latency (e.g. $<$20 ms) after the viewer movement, the so called Motion-to-Photon (MTP) latency constraint \cite{cuervo2018creating_motion_to_photon_latency}. To address the MTP challenge, we consider predictive streaming that predicts the viewer's FoVs for future frames and prefetches tiles within the predicted FoV~\cite{park2019rate,qian_han2020vivo,groot2020}. 

%A smaller tile enables finer granularity for FoV-based pruning, but substantially reduces the coding efficiency because: 1) each tile octree is coded independently with the context for entropy coding confined within the same tile tree, leading to low coding efficiency; 2) the tile root locations are coded independently within the slice header, and the header overhead dominates the slices in the first few levels; 3) grouping the first few levels into a single slice can reduce the header overhead, but it will further reduce the number of spatial layers.

%The Quality-of-Experience (QoE) of PCV viewers is determined by many factors, including the rendered frame quality and quality variations,  streaming continuity and latency, and responsiveness to viewpoint changes, etc.

\section{Progressive PCV Streaming}
Due to the close interaction with PCV objects, viewers are highly sensitive to  QoE impairments, such as black screen, freezes, restarts, excessively long latency, etc. Not only the network and viewer dynamics have direct QoE impacts, bandwidth and FoV prediction errors are also critical for the  effectiveness of predictive streaming. We propose a novel {\it progressive FoV-adaptive PCV  streaming design} to minimize the occurrence of the above impairments and deliver a high level of viewer QoE. 
%in the face of the network and viewer dynamics as well as the errors of predicting them.
%In a typical video streaming session, once the playback starts on the client side, all the video frames should be sequentially rendered and displayed for smooth streaming. 
\subsection{Progressive Downloading/Patching}

\begin{table}[htbp]\caption{Key Variables of Progressive Downloading}
\centering % to have the caption near the table
\begin{tabular}{r c p{6.1cm} }
\toprule
Notation && Meaning \\ 
\hline
$\tau$  & & Download time\\
$t$ & & Playback deadline\\
$T$ & & Video duration\\
$\Delta$ && Download frequency\\
$r_i(t,k)$ & & Downloaded rate in round $t-i$ for tile $k$\\
$f_{ang}$ & & Angular resolution\\
$H$ & & Level of detail (LoD)\\
$Q$ & & Tile quality\\
$B$ & & Bandwidth constraint\\
$R$ & & Maximum rate of tile\\
$\theta$ & & Viewer's span-in-degree across a tile\\
% $\textit{wid}$ & & Physical side length of tile\\
$d$ & & Distance between user viewpoint and tile\\
$\Tilde{p}$ & & Predicted viewing probability \\
\bottomrule
\end{tabular}
\label{tab:notation}
\end{table}

Most of the existing FoV-adaptive streaming solutions can be categorized as  {\it Sequential-Decision-Making (SDM)}: at some time point $\tau$ before the playback deadline $t$ of a frame, one predicts the viewer FoV at $t$,  downloads tiles falling into the predicted FoV at video rates determined by some rate adaptation algorithm, then repeats the process for the next frame. $t-\tau$ is the time interval for both FoV prediction and frame pre-feteching, which is upper bounded by the client side video buffer length. To achieve smooth streaming, a long pre-fetching interval is preferred to absorb the negative impact of  bandwidth variations.  
However, FoV prediction accuracy decays significantly at long prediction intervals. We have studied the optimal trade-off for setting the buffer length for on-demand streaming of 360$^o$ video in~\cite{Liyang_MMSys18,sun2019twotier}.

\begin{figure}[htbp]
\centerline{\includegraphics[width=0.8\linewidth,height=1.6in]{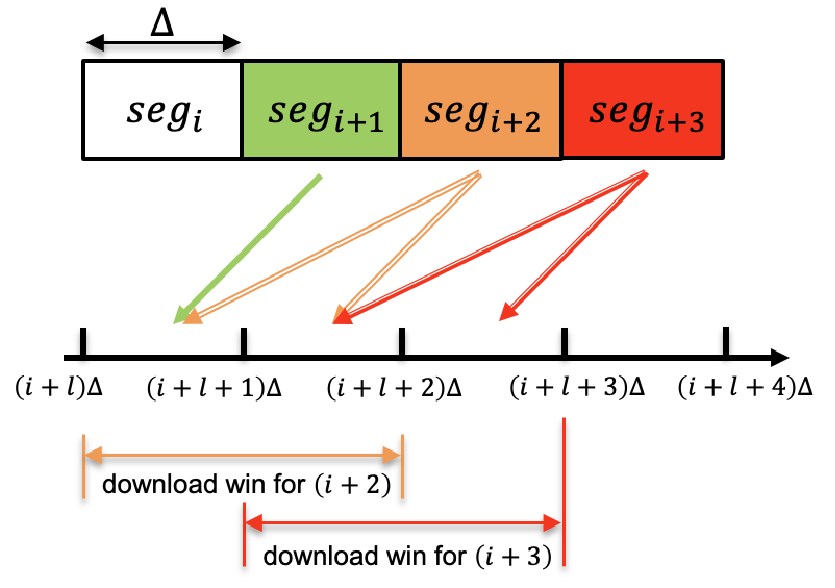}}
\caption{Progressive Streaming Example with Sliding-window Size of $3$: in round $i\Delta$, the base layer of all tiles of  segment $i+3$ within the predicted FoV are being downloaded for the first time, while tiles of segment $i+2$ falling into the predicted FoV are being patched with enhancement layers. Tiles of segment $i+3$ will be patched in the next round $(i+1)\Delta$.}
\label{fig:seg_download}
\end{figure}

 Scalable PCV coding opens up a new dimension to reconcile the conflicting desires for long streaming buffer and short FoV prediction interval. It allows us to {\bf progressively download and patch} tiles: when a frame's playback deadline is still far ahead, we only have a rough FoV estimate for it, and will download  low resolution slices of tiles overlapping with the predicted FoV; as the deadline approaches, we have more accurate FoV prediction, and will {\it patch} the frame by downloading additional enhancement slices for tiles falling into the predicted FoV. {\it Progressive streaming is promising to simultaneously achieve streaming smoothness and robustness against bandwidth variations and FoV prediction errors.} On one hand, as shown in Fig.~\ref{fig:seg_download}, each tile in each segment is downloaded over multiple rounds, the interval of which is $\Delta$. For example, tiles of $seg_{i+3}$ are downloaded in both round $i\Delta$ and round $(i+1)\Delta$. Traditional methods  download the segment only once in round $i\Delta$ based on the FoV prediction at this moment. The benefit of downloading $seg_{i+3}$ in two rounds is that FoV prediction in round $(i+1)\Delta$ is more accurate than in round $i\Delta$. Therefore, the bandwidth can be allocated to the tiles that would be more likely viewed by user. A segment consists of $S$ frames with the total video duration of $\Delta$. Within each round, multiple segments are downloaded simultaneously. And the final rendered quality of each tile is a function of the total downloaded rate (thanks to the scalable coding, there is minimal information redundancy between the multiple progressive downloads). The final rendered quality is less vulnerable to network bandwidth variations and the FoV prediction errors at individual time instants. On the other hand, tile downloading and patching are guided by FoV predictions at multiple time instants with accuracy improving over time. If a tile falls into the predicted FoV in multiple rounds, the likelihood that it will fall into the actual FoV is high, and its quality will be {\it reinforced} by progressive downloading; if a tile only shows up once in the predicted FoV when its playback time is still far ahead, the chance for it to be in the actual FoV is small. Fortunately, the bandwidth wasted in downloading the low-resolution slices  is low.

%Progressive streaming conducts {\bf Parallel-Decision-Making (PDM)}. For a tile $k$ in the frame to be rendered at $t$, we prefetch it in the previous $w$ rounds, from $t-w$ to $t-1$. Let $r_i(t,k)$ denote its download rate in round $t-i$. The final rate for this tile is $\sum_{i=1}^w r_i(t,k)$. Meanwhile, at download time $\tau$, all tiles of frames with playback time from $\tau+1$ to $\tau+w$ are being downloaded/patched, and the total rate is bounded by the available bandwidth, i.e., $\sum_{i=1}^{w} \sum_k r_i(\tau+i,k) \le B(\tau)$. One greedy solution for bandwidth allocation is to maximize the total expected quality enhancements for all the active frames, based on the current FoV estimations for them. Since FoV estimate improves as playback time approaches,  we should assign decreasing weights for frames $\tau+1$ to $\tau+w$ in the objective function.  
\subsection{Optimal Rate Allocation}
\label{sec:optimal}

Progressive streaming conducts {\bf Parallel-Decision-Making (PDM)}. For a tile $k$ in the frame to be rendered at $t$, we prefetch it in the previous $I$ rounds, from $t-I$ to $t-1$. Let $r_i(t,k)$ denote its download rate in round $t-i$. The final rate for this tile is $\sum_{i=1}^I r_i(t,k)$. 
Meanwhile, at download time $\tau$, all tiles of frames with playback time from $\tau+1$ to $\tau+I$ are being downloaded/patched at rates $\{r_i(\tau+i,k), 1 \le i \le I, \forall k\}$, and the total rate is bounded by the available bandwidth $B(\tau)$, i.e., \[\sum_{i=1}^{I} \sum_k r_i(\tau+i,k) \le B(\tau).\] 
Suppose $Q(r,d)$ is the rendered quality of a tile at rate $r$ when viewed from distance $d$, and $\Tilde{p}_{\tau+i,k}$, $\Tilde{d}_{\tau+i,k}$ are the predicted view likelihood and view distance for tile $k$ of frame to be rendered at time $\tau+i$. One greedy solution for bandwidth allocation at download time $\tau$ is to maximize the expected quality enhancements for all the active frames, based on the current FoV estimations:
\begin{align}
\label{eq:progressive}
% & \underset{\{r_i(\tau+i,k)\}} {\mathbf{max}} \sum_{i=1}^{I} \sum_k w_i \Tilde{p}_{\tau+i,k} \nonumber \\ 
% & \cdot \left \{Q\left(\sum_{j=i}^I r_j(\tau+i,k),\Tilde{d}_{\tau+i,k}\right) -Q\left(\sum_{j=i+1}^{I}r_j(\tau+i,k), \Tilde{d}_{\tau+i,k}\right)\right\},
& \underset{\{r_i(\tau+i,k)\}} {\mathbf{max}} \sum_{i=1}^{I} \sum_k w_i \Tilde{p}_{\tau+i,k} \cdot \left \{Q\left(\sum_{j=i}^I r_j(\tau+i,k),\Tilde{d}_{\tau+i,k}\right) \right. \nonumber \\ 
& \qquad \left. -Q\left(\sum_{j=i+1}^{I}r_j(\tau+i,k), \Tilde{d}_{\tau+i,k}\right)\right\},
\end{align}
, where $w_i$ is a  decreasing function of $i$, considering that FoV estimate accuracy drops for far away frames. 
% {\bf ??? what is intention to include the following sentence ??? for ilqr}
% Meanwhile, tile patching at time $\tau$ operates on top of the layers downloaded by the end of $\tau-1$, bandwidth allocation at time $\tau$ should also consider its impacts for the quality enhancements in the future rounds, leading to a stochastic optimal control problem.

\subsection{View-distance based Tile Utility Model}
\label{sec:qr_model_tile}
\begin{figure}[htbp]
\centerline{\includegraphics[width=\linewidth]{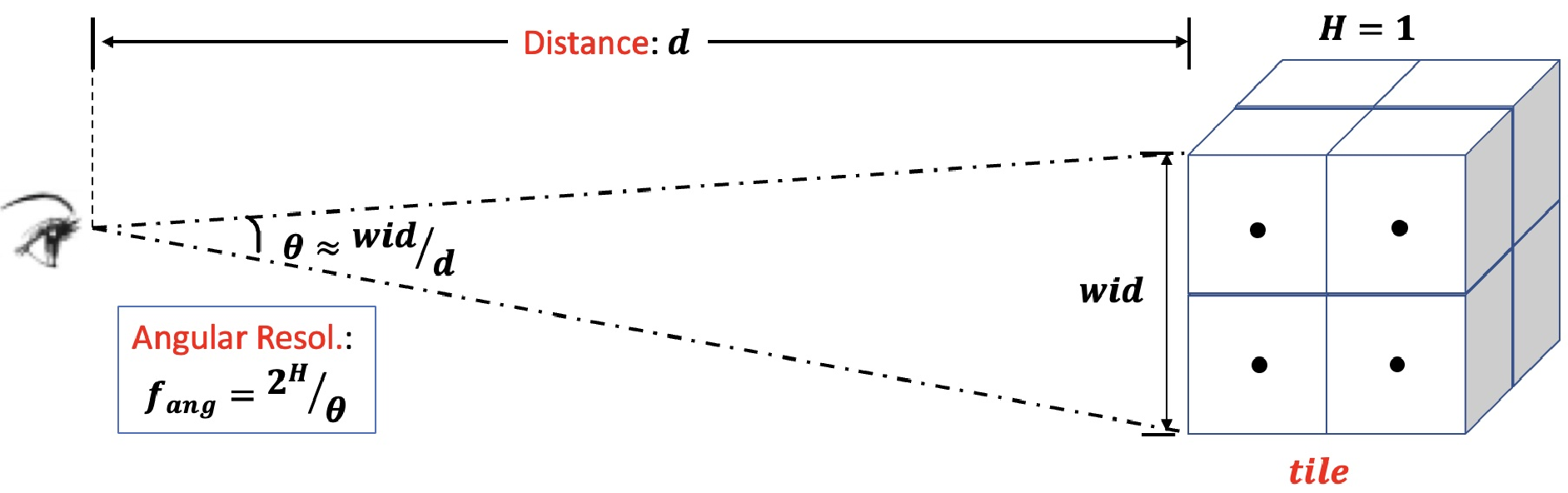}}
\caption{Tile Angular Resolution depends on distance between the viewer and the tile, and  suboctree's height within the tile.}
\label{fig:ang_resol}
\end{figure}

In this section, we explain in detail the proposed tile utility model $Q(r,d)$ in Equation~(\ref{eq:progressive}). Tile quality depends on its angular resolution $f_{ang}$, which is the number of points per degree that are visible to the user within the tile. There are many subjective studies about the quality of rendered images or videos, but most of them only consider the impact of rate, while very few consider the distance between the viewer and the object, which can be very dynamic in PCV streaming. A subjective study in~\cite{qian_han2020vivo} suggests that  users' QoE changes significantly when they view a rendered point cloud object at different distances. The authors showed a reasonable curve telling the relations between utility, rate and viewing distance. Unfortunately, it didn't provide a specific utility model. In our work, we try to find a specific tile utility function with respect to viewing distance and tile rate that generates the same trend of curve in~\cite{qian_han2020vivo}. For the traditional 2D video frames/tiles, which is assumed to be viewed from a fixed distance,  we can infer the direct mapping from rate to quality, which is usually a logarithm function. However, it is not that clear how the rate and viewing distance simultaneously impact the perceived quality of point cloud tiles. A quality model of image with respect to both distance and resolution was proposed in~\cite{westerink1989subjective}, but it doesn't directly fit with point cloud video. Nevertheless, it is inspired from~\cite{westerink1989subjective} that the perceived quality for an object depends on the angular resolution, which depends on the viewing distance and physical size of the object,  as well as image resolution. Following~\cite{westerink1989subjective}, we assume that the perceived quality of a viewed tile is logarithmically increasing with the angular resolution $f_{ang}$, which is the number of points per degree within the tile. With more and more points inside the tile, the per-degree utility increases but more and more slowly. Let $log(c\cdot f_{ang})$ be the per-degree quality of a tile, we assume that the perceived quality of a tile that covers $\theta$ degree in either horizontal or vertical direction is  
\[Q(f_{ang}, \theta)=\theta \cdot \log(c\cdot f_{ang}),\]
where $c$ is a constant factor to be determined by the saturation threshold. As shown in Fig.~\ref{fig:ang_resol}, each tile is a suboctree, and $f_{ang}$ is decided  by the level of detail (LoD) which is the suboctree's height $H$ within the tile, and the viewer's span-in-degree $\theta$ across the tile:
\[f_{ang}=\frac{2^H}{\theta}=\frac{d\cdot 2^H\cdot\pi}{wid\cdot 180},\]
where $d$ is the distance between the viewer and the tile, and $wid$ is the physical side length of a tile. Therefore, the tile utility model is:
\begin{equation}
Q(H,d)=\frac{M}{d} \cdot \log(c\cdot \frac{d\cdot 2^H}{M}),
\label{eq:Q_angle}
\end{equation}
where $M=wid\cdot 180/\pi$.

To achieve a level of detail (LoD) $H$, all nodes up to level $H$ of the tile suboctree has to be delivered, which will incur a certain rate (the total number of bits to code a tile to level $H$). We have coded several test PCVs using MPEG G-PCC and found that the LoD is logarithmically  related to the rate.  Due to tile diversity, the data rate needed to code octree nodes up to level $H$ is highly tile-dependent. In other words, given a coding rate budget $r$, the achievable LoD is tile-dependent. Let $H_{i,k}(r)$ be the rate-LoD mapping for tile $k$ within frame $i$, we assume 
%. In this work, we fit the mapping for every coded tile and each tile maintains a unique set of parameters:
\[H_{i,k}(r)=a_{i,k}\log(b_{i,k}r+1),\]
where the parameters $a_{i,k}, b_{i,k}$ are obtained by fitting the actual rate data for the tiles coded using  G-PCC. %, and it ensures $L_{i,k}=0$ at $r=0$. Thus,
Substituting this expression into (\ref{eq:Q_angle}) yields:
\begin{equation}
\label{eq:qr_model}
Q_{i,k}(r,d)=\frac{M}{d} \cdot \left(a_{i,k}\log2\cdot \log(b_{i,k}\cdot r+1)+\log(\frac{c\cdot d}{M})\right).
\end{equation}
Note that tile utility is not necessarily $0$ at $r=0$, because there is possibly one point inside the tile. There's only one parameter to be decided: $c$, which controls the shape of the curve. We decide its value by reasoning that the utility should saturate  at some point even if we keep increasing rate and/or distance. This is due to human visual acuity: generally we cannot resolve any two points denser than $60$ points per degree~\cite{wiki:visual_acuity}. In other words, the derivative of $Q_{i,k}$ with respect to $d$ should be zero given any rate when $d$ is so large that the angular resolution $f_{ang}$ reaches $60/^\circ$:
\[\frac{\partial Q}{\partial d}=\frac{M}{d^2}\left(1-\log(c\cdot f_{ang}) \right)\]
The derivative goes to zero when $f_{ang}$ is $60/^\circ$ if $c=\frac{e}{60}$.

\begin{figure}[tbp]
\centering{
\subfloat[\footnotesize Utility-Distance curve at different tile rates.]
{\includegraphics[width=0.48\linewidth,height=1.5in]{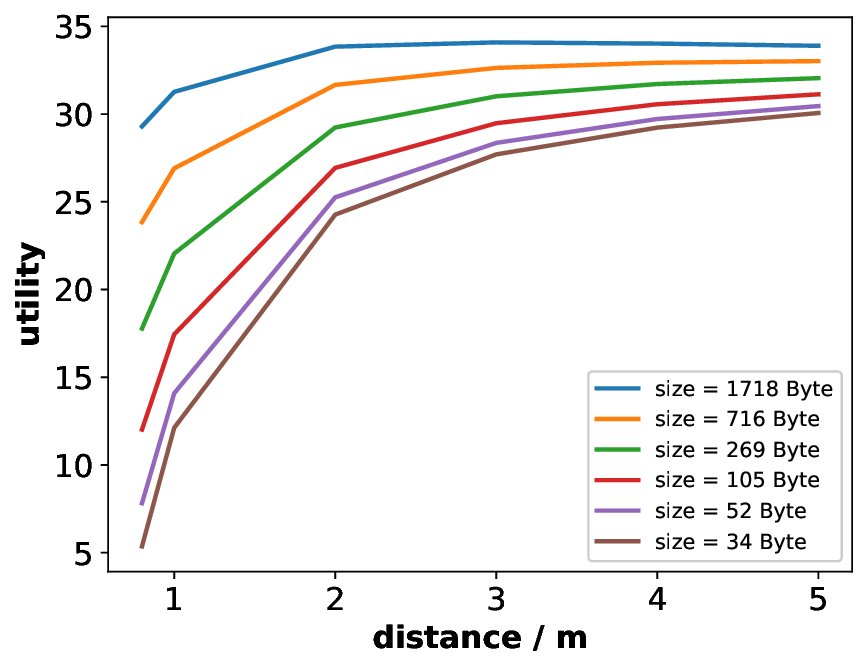}
\label{fig:tile_utility_dist}}
\subfloat[\footnotesize Utility-Rate curve at different distances.]
{\includegraphics[width=0.48\linewidth,height=1.5in]{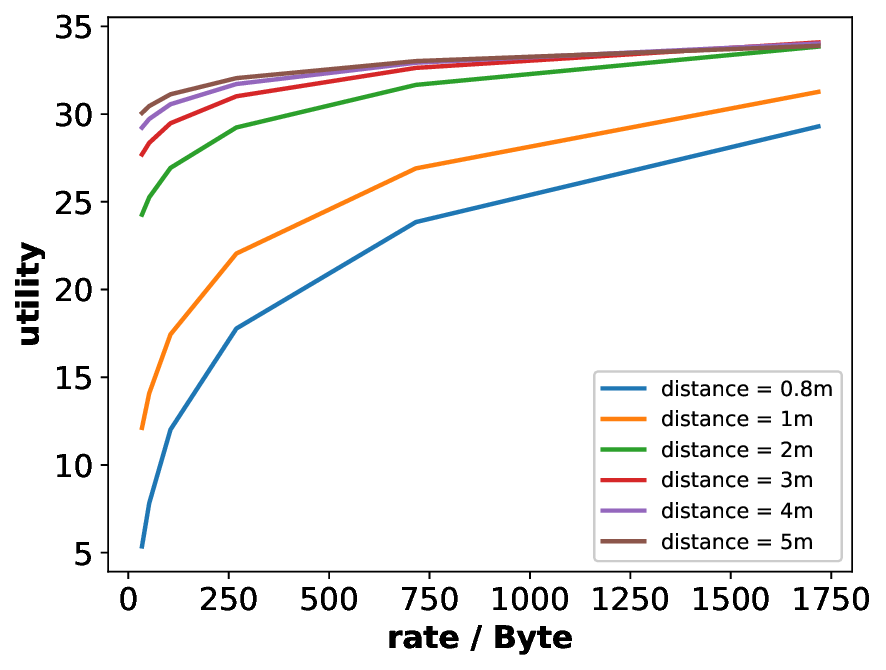}
\label{fig:tile_utility_rate}
}
\caption{Tile Utility Curve.} \label{fig:tile_util_curves}}
\end{figure}

Then we can get the curve similar to the one in~\cite{qian_han2020vivo}. Taking a tile for example, the utility curve with respect to tile rate and distance is shown in Fig.~\ref{fig:tile_util_curves}. It's a tile from point cloud video \textit{Long Dress} of 8i~\cite{subramanyam2020user}, where the figure is $1.8m$ high and tile is at the 4th level of the whole octree. For more details about the dataset, please refer to Section~\ref{sec:evaluation}. For simplicity, we ignore the subscripts of $Q_{i,k}$ as $Q$ in the following sections.
% {\bf ??? you should mention where you got  the curves (or the trend of the curves) in Fig.4, how you used them to get the parameters in (2)???}

% \begin{figure}[htbp]
% \centerline{\includegraphics[width=.8\linewidth]{figs/utility_distance0.eps}}
% % \caption{Tile Utility Curve}
% \label{fig:tile_utility_dist}
% \end{figure}
% \begin{figure}[htbp]
% \centerline{\includegraphics[width=.8\linewidth]{figs/utility_rate0.eps}}
% \caption{Tile Utility Curve}
% \label{fig:tile_utility_rate}
% \end{figure}

\vspace{-.2in}

\subsection{Water-filling Rate Allocation Algorithm}
\label{sec:kkt}
We now develop an analytical algorithm for solving the utility maximization problem formulated in Sec.~\ref{sec:optimal}. In Equation~(\ref{eq:progressive}), the second term  $Q\left(\sum_{j=i+1}^{I}r_j(\tau+i,k), \Tilde{d}_{\tau+i,k}\right)$ represents the tile quality of the layers that have been downloaded to the buffer before the current round, which is a given constant for the rate optimization at the current round. The utility maximization problem for each round is simplified as below, along with the constraints:
\begin{subequations}
\label{eq:neat_progressive}
\begin{align}
\label{eq:neat_progressive_obj}
& \underset{\{r_i(\tau+i,k)\}} {\mathbf{max}} \sum_{i=1}^{I} \sum_k w_i \Tilde{p}_{\tau+i,k} \cdot Q\left(\sum_{j=i}^I r_j(\tau+i,k),\Tilde{d}_{\tau+i,k}\right) \\
& \text{\it subject to} \notag \\
& \label{KKT:C1} r_i(\tau+i,k)\ge 0, \quad 1 \le \tau \le T+I-1, 1 \le i \le I, \forall k,\\
   & \label{KKT:C2} \sum_{j=i}^I r_j(\tau+i,k) \le R(\tau +i,k), \quad 1 \le \tau \le T+I-1, \nonumber \\
   & \qquad \qquad \qquad \qquad \qquad \qquad \qquad 1 \le i \le I, \forall k,\\
   & \label{KKT:C3} \sum_{i=1}^{I} \sum_k r_i(\tau+i,k) \le B(\tau), \quad 1 \le \tau \le T+I-1,
\end{align}
\end{subequations}
where $T$ is the video length, and $R(\tau+i,k)$ is the maximum rate of each tile $(\tau+i,k)$. The tile rate allocation problem is similar to the typical water-filling problem, where we ``fill'' the tiles in an optimal order based on their significance determined by the existing rates, probability to be viewed, distance from the viewer, and the calibration weights $\{w_i\}$.

Equation~(\ref{eq:neat_progressive}) is a nonlinear optimization problem so that the Karush–Kuhn–Tucker (KKT) conditions serve as the first-order necessary conditions. Furthermore, the KKT conditions for this problem are also sufficient due to the fact that the tile utility model $Q(r,d)$ is concave in terms of $r$ with a given $d$,   and the inequality constraints~(\ref{KKT:C1}) and (\ref{KKT:C2}) are continuously differentiable and convex, and (\ref{KKT:C3}) is an affine function.  Therefore, we apply KKT conditions to optimally solve this tile rate allocation problem as shown in Algorithm~\ref{alg:kkt_alloc}. And the detailed KKT condition-based optimization is explained in Section~\ref{sec:kkt_opt}.

\begin{algorithm}
% {\fontsize{9pt}{9pt}\selectfont
\caption{KKT Condition based Tile Rate Allocation}
\label{alg:kkt_alloc}
  \begin{algorithmic}[1]
 	\STATEx  {\textbf{Input:} update window size $I$, point cloud video, video length $T$, user FoV trace, bandwidth trace, utility coefficients of all tiles.}
        \FOR {$\tau$ in 1:$T+I-1$}
            \STATE {Predict FoV for all the tiles in the update window,}
            \STATEx {\indent \indent get viewing probability of tiles: $\Tilde p_{\tau+i,k}, 1 \le i \le I, \forall k$}
            \STATEx {\indent \indent get viewing distances from user viewpoint to tiles:} 
            \STATEx {\indent \indent $\Tilde d_{\tau+i,k}$,}
            \STATE {Call KKT-Condition based optimization (\textbf{Algorithm \ref{alg:kkt_analy}})};
            \STATE {Allocate tile rates based on the results};
            \STATE {Evaluate quality for frames being watched by user};
        \ENDFOR
	% \STATEx {\textbf{Output:} }
 %        \IF {higher resolution tile $c_{i,h}$ in cache, where $h \geq r$}
	% 			    \STATE {$hit++$}
	% \ELSIF{lower resolution tiles $c_{i,l}$ in cache, where $l < r$}
 %            \IF { $S_T(i,r) > S_T(i,l)$}  
 %                \STATE{replace $c_{i,l}$ with $c_{i,r}$}
 %            \ENDIF
 %        \ELSE
 %            \STATE{add $c_{i,r}$ in cache}
 %        \ENDIF
 %        \WHILE{total tile size in cache $>$ Cache Size}        
 %            \STATE{get the lowest $S_T(j,k)$ from cache}
 %            \STATE{remove $c_{j,k}$ from cache}
 %        \ENDWHILE     
\end{algorithmic}
\end{algorithm}

\subsection{KKT Condition-based Optimization}
\label{sec:kkt_opt}
In this section, we explain how to embed KKT condition-based optimization into Problem~\ref{eq:neat_progressive}. We follow the standard procedures of KKT conditions.

\subsubsection{Standard Formulation}
\label{sec:standard_kkt}
First, we need to simplify the symbols in Problem~(\ref{eq:neat_progressive}). At each update time $\tau$, we call this KKT optimization, so we eliminate $\tau$ to simplify the variables: $\sum_{j=i}^I r_i(\tau + i, k)$ is replaced by $r_{i,k}$, which is the resulting tile rate after this update round. Also, we simplify the constants $R(\tau+i,k)$ as $R_{i,k}$, $\Tilde{p}_{\tau+i,k}$ as $\Tilde{p}_{i,k}$, and $\Tilde{d}_{\tau+i,k}$ as $\Tilde{d}_{i,k}$, where $1\le i\le I$ and $k$ is tile index within each frame $i$. What's more, we define $r_{i,k}^0 = \sum_{j=i+1}^I r_j(\tau+i,k)$, the existing downloaded rate for tile $(\tau+i,k)$. Problem~\ref{eq:neat_progressive} can be simplified as below:
\begin{subequations}
\label{eq:simplify_neat_progressive}
\begin{align}
\label{eq:simplify_neat_progressive_obj}
& \underset{\{r_{i,k}\}} {\mathbf{max}} \sum_{i,k} w_i \Tilde{p}_{i,k} \cdot Q\left(r_{i,k},\Tilde{d}_{i,k}\right) \\
\text{\it subject to} \notag \\
& \label{KKT:simplify_C1} r_{i,k}\ge r_{i,k}^0, \quad 1 \le i \le I, \forall k,\\
   & \label{KKT:simplify_C2} r_{i,k} \le R_{i,k}, \quad \forall 1 \le i \le I, \forall k,\\
   & \label{KKT:simplify_C3} \sum_{i,k} (r_{i,k} - r^0_{i,k}) \le B,\quad \forall 1 \le i \le I, \forall k,
\end{align}
\end{subequations}

Next, we define several functions for different parts of Problem~(\ref{eq:neat_progressive}). Let $f(\textbf{r})=\sum_{i,k} w_i \Tilde{p}_{i,k} \cdot Q\left(r_{i,k},\Tilde{d}_{i,k}\right)$, $\mathbf{g}^1(\mathbf{r})=-\mathbf{r}+\mathbf{r^0}$, $\mathbf{g}^2(\mathbf{r})=\mathbf{r}-\mathbf{R}$, and $h(\mathbf{r})=\sum_{i,k} (r_{i,k} - r^0_{i,k}) - B$, where $\textbf{r}$, $\mathbf{r^0}$ and $\mathbf{R}$ are vectors of $r_{i,k}$, $r^0_{i,k}$ and $R_{i,k}$ respectively. Then we rewrite the above equations once more into the standard form of KKT condition:
\begin{subequations}
\label{eq:standard_neat_progressive}
\begin{align}
\label{eq:standard_neat_progressive_obj}
& \underset{\mathbf{r}} {\mathbf{min}} -f(\mathbf{r}) \\
\text{\it subject to} \notag \\
& \label{KKT:standard_C1} \mathbf{g}^1(\mathbf{r}) \le 0, \\
   & \label{KKT:standard_C2} \mathbf{g}^2(\mathbf{r}) \le 0, \\
   & \label{KKT:standard_C3} h(\mathbf{r})=0,
\end{align}
\end{subequations}

\subsubsection{Analytical Algorithm}
The solutions satisfying KKT conditions for this problem are also sufficient because $f(\cdot)$ is concave, both $\mathbf{g}^1(\cdot)$ and $\mathbf{g}^2(\mathbf{\cdot})$ are continuously differentiable and convex, and $h(\cdot)$ is affine. We follow the standard steps to solve it. With Lagrange multipliers, (\ref{eq:standard_neat_progressive}) is transformed as:
\begin{subequations}
\label{eq:lagrange}
\begin{align}
\underset{\boldsymbol{\mu^1,\mu^2},\lambda} {\mathbf{max}} \quad \underset{\mathbf{r}} {\mathbf{min}} \quad & L(\mathbf{r},\boldsymbol{\mu}^1,\boldsymbol{\mu}^2,\lambda)=-f(\mathbf{r})\\
&+\boldsymbol{\mu^1\cdot \mathbf{g}^1(\mathbf{r})}+\boldsymbol{\mu^2\cdot \mathbf{g}^2(\mathbf{r})} +\lambda h(\mathbf{r}).
\end{align}
\end{subequations}
Then we derive the properties of the solution based on the following KKT conditions:

\textbf{Stationarity}:
\begin{equation}
\label{eq:stationarity}
\frac{\partial L}{\partial \mathbf{r}}=0,
\end{equation}
therefore, 
%\[-\frac{z_k}{r_k+1/b_k}-\mu_k^1+\mu_k^2+\lambda=0\]
\[r_k=\frac{z_k}{\lambda+\mu_k^2-\mu_k^1}-\frac{1}{b_k},\]
where $z_k=a_k\theta_k\log2$ is a constant when $\theta_k$ is given.
% \textbf{Primal feasibility}:
% \begin{subequations}
% \label{eq:primal}
% \begin{align}
% \mathbf{g}^i(\mathbf{r})\le 0 \\
% h(\mathbf{r})=0,
% \end{align}
% \end{subequations}

\textbf{Dual feasibility}:
\begin{equation}
\label{eq:dual_cond}
% \begin{align}
\mu_k^i\ge 0, \quad i=\{1,2\},
% \end{align}
\end{equation}

\textbf{Complementary slackness}:
\begin{equation}
\label{eq:slack_cond}
\boldsymbol{\mu}^i\cdot \mathbf{g}^i(\mathbf{r})=0, \quad i=\{1,2\},
\end{equation}
therefore,
\[\mu_k^1(-r_k+r^0_k)=0\]
\[\mu_k^2(r_k-R_k)=0\]

Therefore, the optimal rate allocation has three possible solutions according to conditions~(\ref{eq:stationarity}) and (\ref{eq:slack_cond}):
\begin{equation}
\label{eq:rate_cases}
% &r_k^*=r_k^0, &if \lambda\ge\frac{z_k}{r_k^0+1/b_k};\\
% &r_k^*=R_k,  &if \lambda\le\frac{z_k}{R_k+1/b_k};\\
% &r_k^*=\frac{z_k}{\lambda}-\frac{1}{b_k}, &if \lambda=\frac{z_k}{r_k^*+1/b_k}.
r_k^*=
\begin{cases}
      r_k^0 & \text{if $\lambda\ge\frac{z_k}{r_k^0+1/b_k}$;}\\
      R_k & \text{if $\lambda\le\frac{z_k}{R_k+1/b_k}$;}\\
      \frac{z_k}{\lambda}-\frac{1}{b_k} & \text{if $\lambda=\frac{z_k}{r_k^*+1/b_k}$.}
    \end{cases}     
\end{equation}

Once $\lambda$ is decided, all $r_k's$ are decided. Therefore, the goal is to find the optimal $\lambda$. The analytical water filling algorithm for finding $\lambda^*$ is described in Algorithm~\ref{alg:kkt_analy}.

\begin{algorithm}
% {\fontsize{9pt}{9pt}\selectfont
\caption{Analytical Water Filling Algorithm}
\label{alg:kkt_analy}
  \begin{algorithmic}[1]
 	\STATEx  {\textbf{Input:} given $\theta_k$ and utility parameters $a_k$ and $b_k$, each tile $k$ keeps two values: $V_k^1=\frac{z_k}{r_k^0+1/b_k}$ and $V_k^2=\frac{z_k}{R_k+1/b_k}$.}
        \FOR {$\lambda$ in [0\quad :\quad $max\{V_k^1\}$]}
            \STATE {compare $\lambda$ to both $V_k^1$ and $V_k^2$, $\forall k$,}
            \STATEx {\indent \indent obtain resulted $r_k$ based on Equation (\ref{eq:rate_cases}).}
            \IF{$\sum_k (r_k-r_k^0)=B$}
                \State $\lambda^*=\lambda$,
                \State $r_k^*=r_k,\quad \forall k$,
                \State break.
            \ENDIF
        \ENDFOR
\end{algorithmic}
\end{algorithm}

\subsubsection{Complexity Analysis}
\label{sec:complexity}
From Algorithm \ref{alg:kkt_alloc}, at each update time $\tau$, there are three steps to output the tile rate allocation solution.
\begin{enumerate}
    \item \textbf{FoV prediction}: We use truncated linear regression with history window size $0.5\cdot I$ to predict future $I$ viewpoints. The regression time in worst case is $O(I^2)$ using \textit{sklearn} package in Python. The inference time is $O(I)$.
    \item \textbf{Hidden Point Removal (HPR)}: We use the HPR function in open3d package in Python. The time complexity is $O(nlogn)$ using the algorithm in~\cite{katz2007direct}, where $n$ is maximum number of points in each frame. To make HPR faster, we downsampled the original point cloud to just $n\approx 10k$ points per frame before running HPR.
    \item \textbf{KKT-Condition based optimization}: Algorithm~\ref{alg:kkt_analy} performs a line search looking for the optimal $\lambda$. The line search takes at least $I$ steps  in the range $[0, max\{V_k^1\}]$. And within each loop we need to obtain the rates for all the tiles within the update window, the complexity of which is $O(\alpha\cdot I)$ where $\alpha$ is the number of tiles in a frame that is around $50$ on average. Thus, the time complexity of line search is $O(\alpha^2I^2)$.\\
    
    However, we actually implement Algorithm~\ref{alg:kkt_analy} using binary search, which is obviously faster than line search. Before binary search, we sort the values $V_k$ which costs $O(\alpha I\cdot log(\alpha I))$. Then, similar to the analysis of line search, the binary search process costs $O(\alpha I\cdot log(\alpha I))$, which has the same complexity as the sorting process.\\
\end{enumerate}
To sum up, the time complexity at each update time $\tau$ is $O(I^2+\alpha I\cdot log(\alpha I)+nlogn)$, where $I\le 600$, $\alpha\approx 50$ and $n\approx 10k$. All the variables, $I, n, \alpha$, are relatively small, so the algorithm can achieve the update rate of $30fps$ on moderately configured server with Intel(R) Xeon(R) Platinum 8268 CPU @ 2.90GHz.

\section{Evaluation}
\label{sec:evaluation}

In this section, we demonstrate the superiority of the proposed approach over several baselines.

\subsection{Setup}
\label{sec:exp_setup}
\subsubsection{Datasets}
We used point cloud videos  from the 8i website~\cite{d20178i}, which includes 4 videos, each with $300$ frames with frame rate of $30$ fps. The user FoV trace is from \cite{subramanyam2020user} which involves $26$ participants watching looped frames of the 8i videos on Unity game engine. The bandwidth trace is from NYU Metropolitan Mobile Bandwidth Trace~\cite{mei2019realtime}, which includes 4G bandwidth traces collected in NYC when traveling on buses, subways, or ferry.

\subsubsection{Key Parameters}
\label{sec:exp_config}
In this section, we show the table of the key parameters used in our experiment.

\begin{table}[htbp]\caption{Key Configuration Parameters}
\centering % to have the caption near the table
\begin{tabular}{r c p{1.5cm} }
\toprule
Parameter && Value \\ 
\hline
Download frequency $\Delta$  & & $1s$ \\
Update window size $I$ & & $20s$\\
Segment length & & $1s$\\
Frame rate & & $30fps$\\
number of tile rate versions && $6$\\
FoV prediction history win & & $\le 10s$\\
FoV span & & $90^{\circ}$\\
Bandwidth prediction history win & & $5s$\\
\bottomrule
\end{tabular}
\label{tab:param_config}
\end{table}

We set update window size (buffer length) $I=20s$, which is a relatively long buffer, for unlike the traditional 2-D video, point cloud video is more sensitive to the bandwidth variations due to its higher data rate, so a longer buffer/larger interval is better at handling unstable bandwidth condition. Nowadays, even for 2-D videos, long video buffers are common in Netflix or Youtube VoD streaming. In addition, the larger the interval, the less accurate the FoV prediction. In our case, it's extremely hard to get accurate FoV prediction for $20s$ ahead, and that's why non-progressive method doesn't work well. Nonetheless, the proposed KKT-based progressive framework performs well even if the interval is as large as $20s$, because it progressively downloads every frame over  multiple rounds as the frame is moving toward buffer front and FoV prediction becomes more and more accurate.  As a result, the proposed progressive framework can achieve smooth streaming with long buffers while dealing well with the FoV prediction errors at a large prediction  intervals.

\subsubsection{FoV Prediction and Bandwidth Prediction}
\label{sec:fov_bw_pred}
FoV prediction method used in this work is linear regression, while other approaches like neural network based prediction model can also be used in our proposed framework. More specifically, we predict at most future $I=20s$ simultaneously using past $0.5\cdot I$ frames, which is an empirical best history window size reported in~\cite{qian_han2020vivo}. The FoV span is $90^{\circ}$. The prediction accuracy results with different future prediction window sizes are reported in Section~\ref{sec:kkt-const_vs_nonprog}. 

We predict future bandwidth by harmonic mean, which is a common and simple bandwidth prediction method. More specifically, it predicts the bandwidth in future one second by calculating the harmonic mean of previous bandwidth within a history window. In our work, we set the history window size to $5$ seconds. Since bandwidth prediction is orthogonal to the tile rate allocation algorithm design, we didn't use more complicated bandwidth prediction methods.

\subsection{Baselines}
\label{sec:baseline}
\begin{enumerate}
% \item \textbf{Homogeneous Tiles}: Tiles in point cloud video are more heterogeneous compared to 360 video and 2D video, due to the fact that different tiles have different number of points at the highest density level. Therefore, we must consider this tile heterogeneity of utility model when optimizing Equation~\ref{eq:neat_progressive}. We demonstrate this point by comparing with homogeneous-tile experiment where we assume all the tiles have the same coefficients in the utility model.
% \item \textbf{Uniform Distance}: Distance between each tile and the user viewpoint  plays an important role when evaluating the tile utility in PCV streaming, which makes PCV extremely different from 360 video and 2D video. To demonstrate the importance of distance, in this baseline we assume all the tiles have the same distance from the user viewpoint.
\item \textbf{Non-progressive Downloading}: This baseline is similar to the traditional video streaming in 2D video with one new segment downloaded to the end of the streaming buffer in each round. The rate allocation between frames within each new segment is optimized using a KKT algorithm similar to Algorithm~\ref{alg:kkt_alloc} and~\ref{alg:kkt_analy}, based on just one-time FoV prediction result.
% \item \textbf{Greedy heuristic}: One state-of-art window based tiles rate allocation allocation is proposed in~\cite{park2019rate}, in which authors proposed a greedy heuristic algorithm without scalable PCV coding.
\item \textbf{Progressive with Equal-allocation}: To demonstrate the effectiveness of KKT based optimization approach for progressive streaming, we evaluate a naive progressive streaming baseline where the available bandwidth is equally allocated over all active tiles falling into the predicted FoV at each round. 
% \item \textbf{Distance-unaware utility}: Distance between each tile and the user viewpoint  plays an important role when evaluating the tile utility in PCV streaming, which makes PCV extremely different from 360 video and 2D video. To demonstrate the importance of distance-aware utility model, in this baseline we assume a tile's utility is independent of the viewing distance.
\item \textbf{Rate-Utility Maximization Algorithm (RUMA)}: In the related state-of-the-art work~\cite{park2019rate}, the authors proposed a tile utility model where the tile utility is simply the logarithm of the tile rate multiplied by the number of differentiable points. Based on this model, the authors further proposed a greedy heuristic tile rate allocation algorithm which we call RUMA for short. RUMA didn't explicitly introduce the concept of scalable PCV coding. Different from RUMA, in our work, we formulate the PCV streaming problem with a well developed concept of scalable coding, and propose a more refined utility model that better reflect viewer's visual experience when viewing a point cloud tile from certain distance. To achieve this, we have coded several test PCVs using MPEG G-PCC and found that the level of detail (LoD) is logarithmically related to the rate.  Due to tile diversity, the data rate needed to code octree nodes up to level $H$ is highly tile-dependent. In other words, given a coding rate budget $r$, the achievable LoD is tile-dependent. We fit the coefficients of this logarithmic mapping from rate to LoD, and model the tile utility as a function of the  angular resolution to replicate the empirical  subjective utility-(rate, distance) curve  in~\cite{qian_han2020vivo}, which makes our utility model more convincing. Also, we adopt KKT to solve the continuous version of the rate allocation problem then round the rate to its discrete version. We show through simulation experiments that our method outperforms RUMA.

% The tile utility model didn't catch the true relation between the utility and the distance from the user viewpoint. Furthermore, the greedy rate allocation algorithm includes too much approximation about the marginal utility of every tile. Instead, in our work, we detailedly introduced the concept of scalable PCV coding and proposed a more accurate tile utility model. 
\end{enumerate}

% \subsection{Metrics}
% \label{sec:metrics}
% proposed: frame quality per degree, delivered angular resolution, number of pts per degree
% conventional: psnr/ssim

\subsection{Experimental Results}
\label{sec:experiment}
We run simulations  with real users' FoV traces~\cite{subramanyam2020user} over four 8i videos. The height of the point cloud cube is $1.8m$ and roots of tiles are at $4th$ level of the whole octree. Each tile subtree has 6 different levels of detail. The update window size is $I=20s$, and it moves forward every one second. We are using $10$ FoV traces and $3$ bandwidth traces, and the quality results are averaged over frames in Table~\ref{tab:ang_resol} and \ref{tab:frame_quality_per_degree}.

Table~\ref{tab:ang_resol} and \ref{tab:frame_quality_per_degree} display mean and variance of different metrics in different scenarios. When both the bandwidth and the FoV oracles are available (Scenario A), non-progressive streaming is comparable with progressive streaming. But in the realistic setting with network bandwidth estimation errors and FoV prediction errors (Scenario B), our proposed progressive streaming solution with either constant or exponentially decreasing frame weights can deliver much higher quality in most cases, measured using both the \textit{quality per degree} and the \textit{delivered angular resolution} or the number of points per viewing degree, averaged over all FoV tiles.

\begin{table}[htb]%
\centering
    \centering
    % \vspace{-8mm}
    \caption{\small Mean and Variance of Angular-Resolution-per-Frame.}
    {
    \begin{tabular}{c|c|c|c|c|c} \hline
       Scen-  & Non- & Equal- & RUMA & \multicolumn{2}{c}{Progressive KKT}   \\ \cline{5-6} 
       ario & Progressive & Split &  & Constant & Exp \\ \hline 
       % A & 18.97, 3.09 & 17.33, 1.63 & 19.52, 1.71 & 19.50, 1.66 & N/A \\ \hline
       % B & 6.19, 6.06 & 15.09, 2.06 & 14.83, 2.07 & 15.48, 1.85 & 16.87, 1.78 \\  \hline 
       % C &  & 15.20, 2.08 & & \\  \hline 
       A & 18.9, 3.0 & 17.3, 1.6 & 19.5, 1.7 & 19.5, 1.6 & N/A \\ \hline
       B & 6.1, 6.0 & 15.0, 2.0 & 15.9, 2.5 & 15.4, 1.8 & 18.3, 1.9 \\  \hline 
    \end{tabular}
    }
    
    \label{tab:ang_resol}
   % \vspace{-5mm}
\end{table}

% \begin{table*}[htb]%
% \centering
%     \centering
%     % \vspace{-8mm}
%     {\small
%     \begin{tabular}{c|c|c|c|c|c} \hline
%        Test  & Non- & Equal- & Distance- & \multicolumn{2}{c}{Progressive KKT}   \\ \cline{5-6} 
%        Scenario & Progressive & Split & Unaware & Constant & Exp \\ \hline 
%        A & 19.02, 3.03 & 17.33, 1.54 & 19.37, 1.60 & 19.31, 1.56 & N/A \\ \hline
%        B & 6.53, 5.78 & 15.14, 2.07 & 14.83, 2.07 & 15.06, 1.93 & 17.08, 2.01 \\  \hline 
%        % C &  &  & & \\  \hline 
%     \end{tabular}
%     }
%     \caption{\small Average Frame-Angular-Resolution}
%     \label{tab:ang_resol}
%    % \vspace{-5mm}
% \end{table*}

% \begin{table*}[htb]%
% \centering
%     \centering
%     % \vspace{-8mm}
%     {\small
%     \begin{tabular}{c|c|c|c|c|c} \hline
%        Test  & Non- & Equal- & Distance- & \multicolumn{2}{c}{Progressive KKT}   \\ \cline{5-6} 
%        Scenario & Progressive & Split & Unaware & Constant & Exp \\ \hline 
%        A & -61.48, 57.98 & -86.33, 36.01 & -48.31, 29.75 & -46.04, 28.10 & N/A \\ \hline
%        B & -806.90, 451.14 & -134.13, 59.25 & -131.01, 58.89 & -123.13, 52.25 & -94.40, 46.74 \\  \hline 
%        % C &  &  & & \\  \hline 
%     \end{tabular}
%     }
%     \caption{\small Average Frame Quality}
%     \label{tab:frame_quality}
%    % \vspace{-5mm}
% \end{table*}

\begin{table}[htb]%
\centering
    \centering
    % \vspace{-8mm}
    \caption{\small Mean and Variance of Per-Degree Frame Quality}
    {
    \begin{tabular}{c|c|c|c|c|c} \hline
       Scen-  & Non- & Equal- & RUMA & \multicolumn{2}{c}{Progressive KKT}   \\ \cline{5-6} 
       ario & Progressive & Split &  & Constant & Exp \\ \hline 
       A & 5.45, 0.42 & 5.34, 0.39 & 5.49, 0.39 & 5.50, 0.38 & N/A \\ \hline
       B & 2.50, 1.53 & 5.15, 0.44 & 4.46, 0.63 & 5.16, 0.39 & 5.42, 0.41 \\  \hline 
       % C &  &  & & \\  \hline 
    \end{tabular}
    }
    
    \label{tab:frame_quality_per_degree}
   % \vspace{-5mm}
\end{table}

\begin{table}[htb]%
\centering
    \centering
    % \vspace{-8mm}
    \caption{\small Average Per-frame Wasted Bandwidth: bandwidth used to download tiles that are not in the user's actual FoV.}
    {
    \begin{tabular}{c|c|c|c|c} \hline
       Scen-  & Non- & Equal- &  \multicolumn{2}{c}{Progressive KKT}   \\ \cline{4-5} 
       ario & Progressive & Split &  Constant & Exp \\ \hline 
       % A & -61.48, 57.98 & -86.33, 36.01 &  & -46.04, 28.10 & N/A \\ \hline
       B & 13.49 KB & 6.34 KB &  11.78 KB & 5.19 KB \\  \hline 
       % C &  &  & & \\  \hline 
    \end{tabular}
    }
    
    \label{tab:wasted_rates}
   % \vspace{-5mm}
\end{table}

We believe that in addition to the average results from Table~\ref{tab:ang_resol} and \ref{tab:frame_quality_per_degree}, it's also very important to show the frame quality evolution across time in specific streaming sessions in Fig. \ref{fig:KKT-const_non-prog} and Fig. \ref{fig:other_videos_kkt-const_non-prog_frame_quality_per_degree_bw5} to demonstrate the drawback of non-progressive baseline as well as the superiority of progressive downloading due to its multi-round downloading with more and more accurate FoV prediction for every frame. In the following sections, we show the comparison results in real-world Scenatio B only. In Section~\ref{sec:kkt-const_vs_nonprog}, we show results for each of the four 8i videos, while the rest of the sections only present the detailed results for \textit{Long Dress} because all the other videos have similar trends.

\subsubsection{\textbf{KKT-const V.S. non-progressive}}~\par
\vskip1mm
\label{sec:kkt-const_vs_nonprog}
 % Fig.~\ref{fig:KKT-const_non-prog} shows results for one 8i video \textit{Long Dress}, and Fig.~\ref{fig:other_videos_kkt-const_non-prog_frame_quality_per_degree_bw5} shows per-degree frame quality comparisons for other three 8i videos.

\begin{figure*}[!t]
    \centering 
    \subfloat[\footnotesize Angular Resolution per Frame]{\includegraphics[width=.3\linewidth]{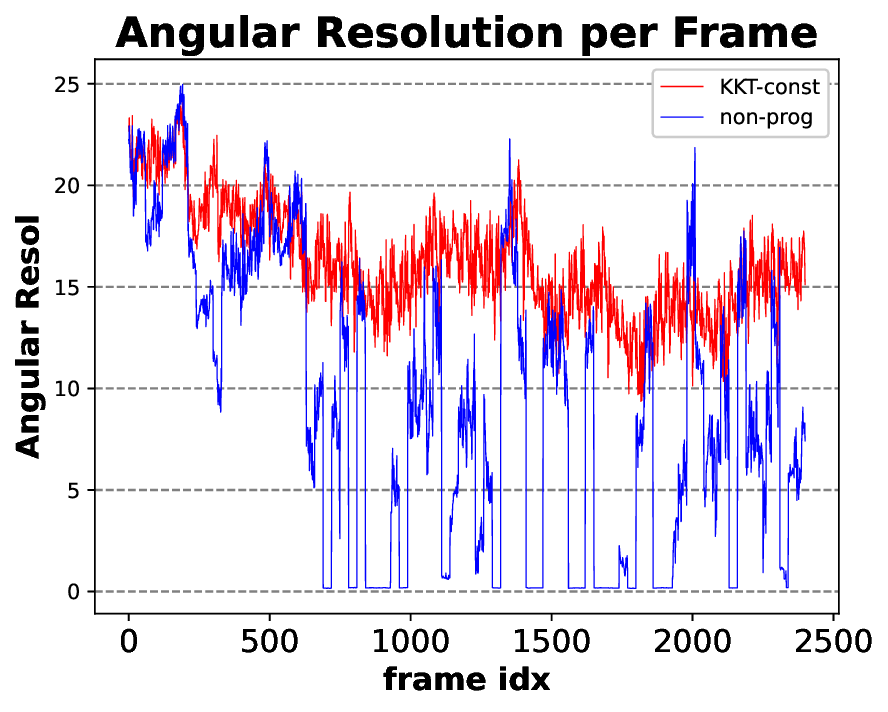}%
    \label{fig:KKT-const_non-prog_ang_resol_bw5}}
    \hfil
    \subfloat[\footnotesize Per-Degree Frame Quality]{\includegraphics[width=.3\linewidth]{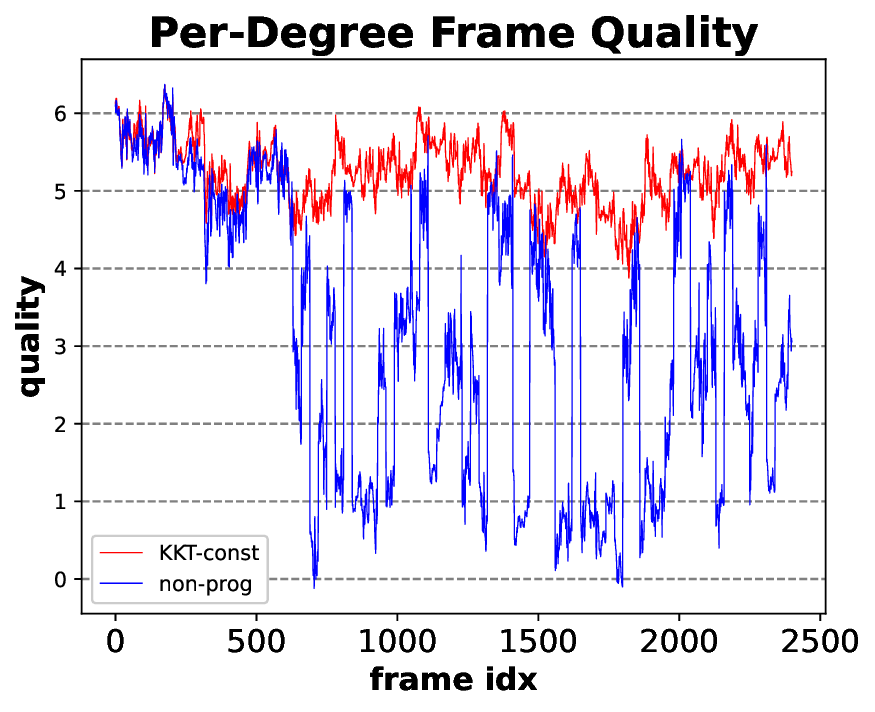}%
    \label{fig:KKT-const_non-prog_frame_quality_per_degree_bw5}}
    \hfil
    \subfloat[\footnotesize Wasted Bandwidth per Frame]{\includegraphics[width=.3\linewidth]{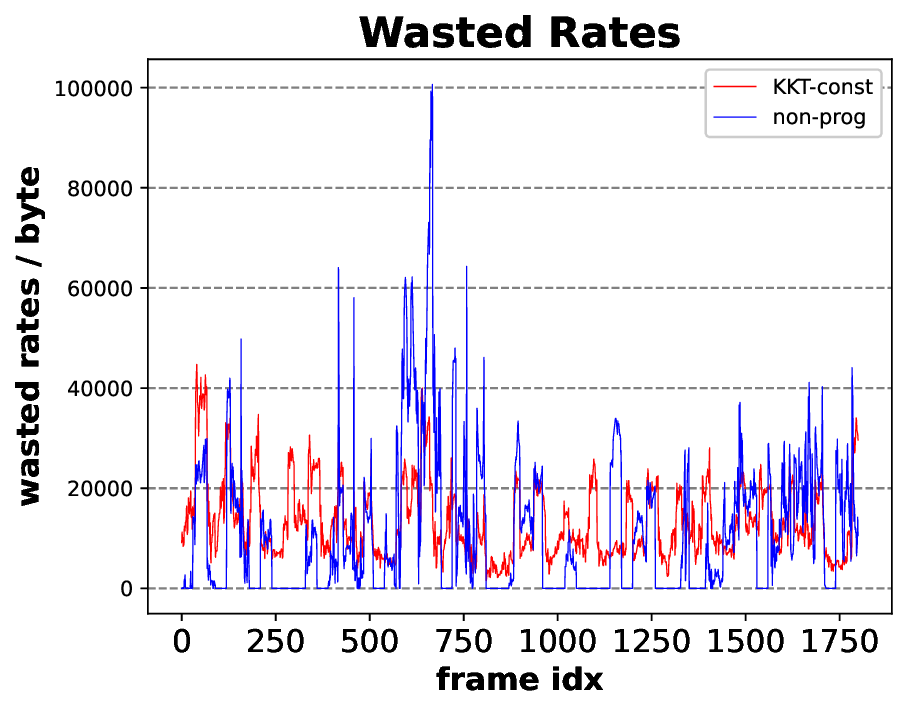}%
    \label{fig:KKT-const_non-prog_wasted_rates_bw5}}
\caption{Compare KKT-const and non-progressive}
\label{fig:KKT-const_non-prog}
\end{figure*}

\begin{figure*}[!t]
    \centering 
    \subfloat[\footnotesize \textit{Soldier}]{\includegraphics[width=.3\linewidth]{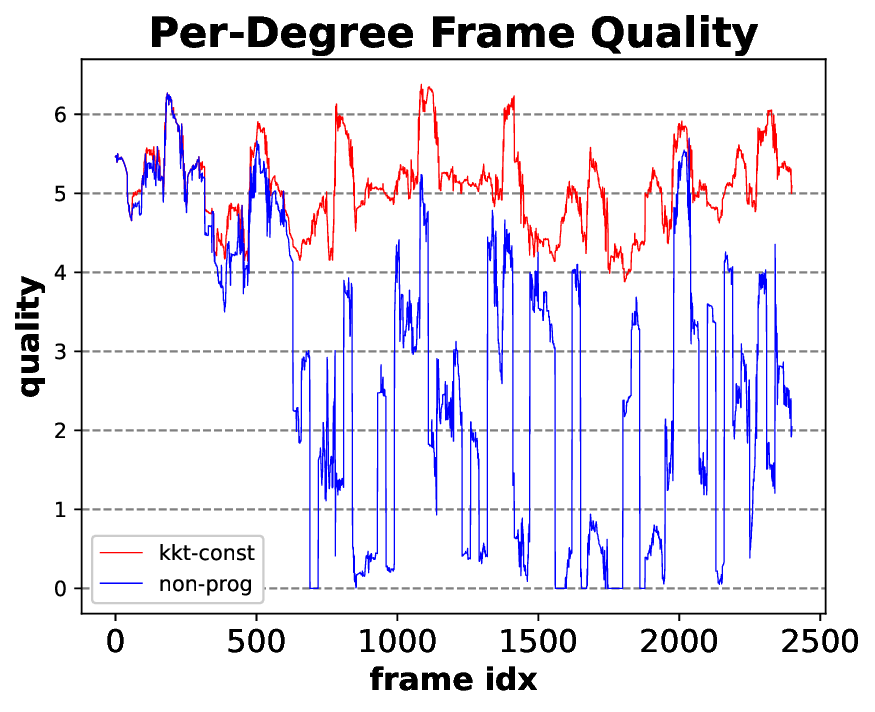}%
    \label{fig:soldier}}
    \hfil
    \subfloat[\footnotesize \textit{Loot}]{\includegraphics[width=.3\linewidth]{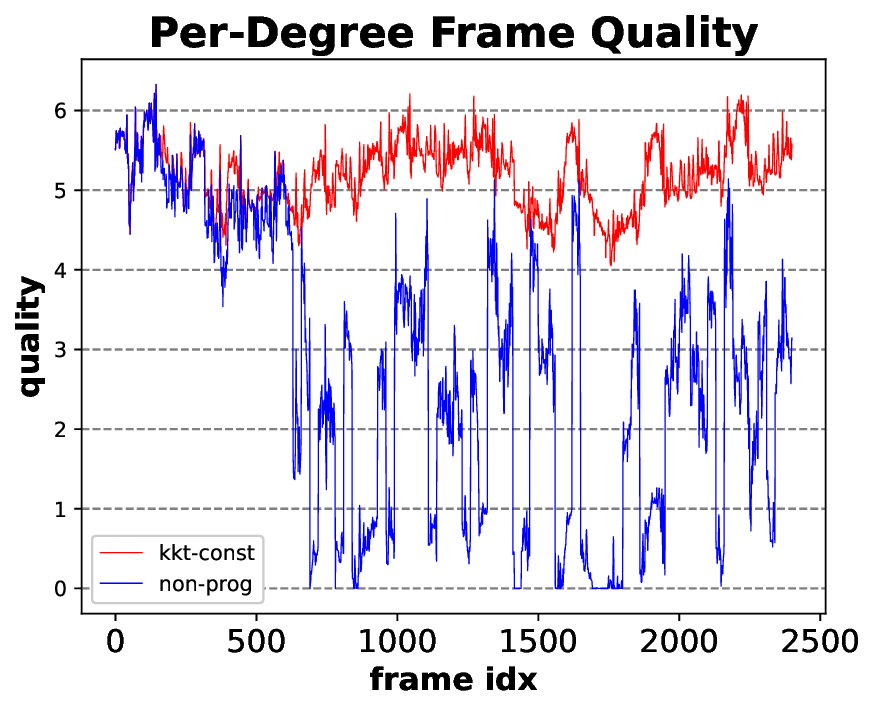}%
    \label{fig:loot}}
    \hfil
    \subfloat[\footnotesize \textit{Red and Black}]{\includegraphics[width=.3\linewidth]{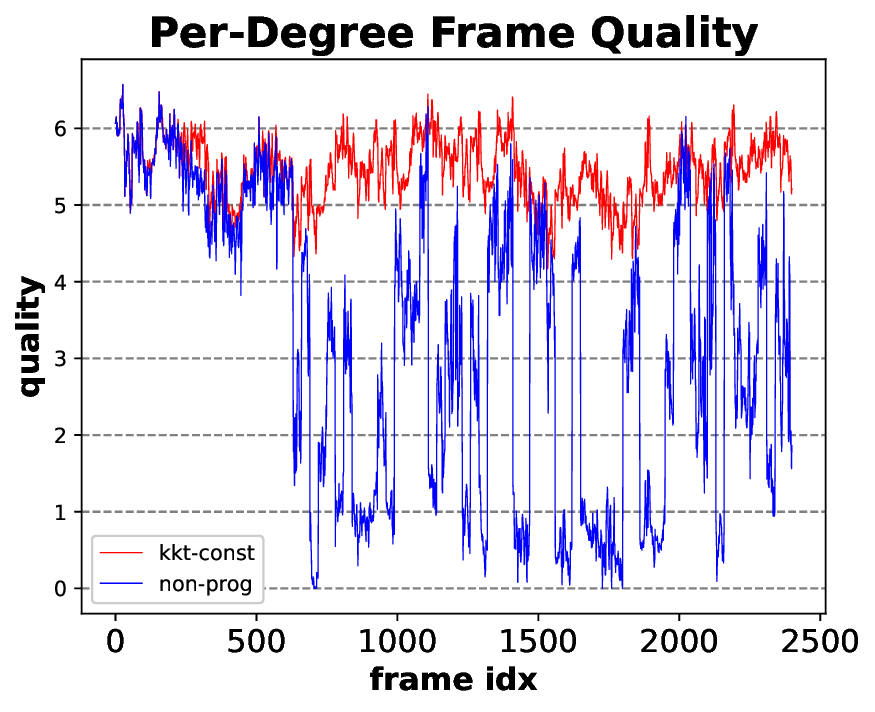}%
    \label{fig:redandblack}}
\caption{Compare KKT-const and non-progressive for other videos.}
\label{fig:other_videos_kkt-const_non-prog_frame_quality_per_degree_bw5}
\end{figure*}

\textbf{Quality Supremacy}: Only the tiles in the user's actual FoV are counted into the final quality using Equation (\ref{eq:Q_angle}). Fig.~\ref{fig:KKT-const_non-prog_ang_resol_bw5} and Fig.~\ref{fig:KKT-const_non-prog_frame_quality_per_degree_bw5} as well as Fig.~\ref{fig:other_videos_kkt-const_non-prog_frame_quality_per_degree_bw5} show the frame quality evolution results. A drop around frame index $600$ can be observed. The reason is that we assume the server knows the initial 6-DoF viewpoint of the user if he/she starts watching the first frame. Since it's cold start, the first $600$ frames will be downloaded before being watched, so we need to predict FoV for each of the $600$ frames and only download the tiles within the predicted FoV. Obviously the predicted 6-DoF viewpoints for the first $600$ frames using Linear Regression (LR) would be the same as the initial 6-DoF viewpoint, because there's only one history data sample which is exactly the initial 6-DoF viewpoint. And it turns out that the constant predicted viewpoint for the first $600$ frames is not too bad because the viewer doesn't move dramatically away from the initial viewpoint. Therefore, the quality of first $600$ frames are not bad. However, after the cold start, when the viewer starts watching more frames, the FoV prediction for the following frames (from $601^{st}$ frame) will be quite different than the initial viewpoint using Linear Regression (LR) because there are more than one history viewpoints in the LR history window now. Since the prediction horizon is as large as future $600^{th}$ frame ($20s$ later), the prediction accuracy would be very bad. Therefore, we observe the drop in the non-progressive curve right after the first $600$ frames in the figures. The results mainly demonstrate that the non-progressive framework is super sensitive to FoV prediction error under a long buffer, while the proposed progressive framework can deal with that by progressively downloading every frame for multiple rounds as the frame is moving toward buffer front and FoV prediction becomes more and more accurate.

Except for the first $600$ frames where the FoV prediction accuracy for both algorithms is accurate, KKT-const, which uses constant weights for all frames in KKT calculation, dominates non-progressive baseline in terms of both angular resolution and per-degree quality by a large gap on average. The reason is that the non-progressive baseline predicts FoV for every frame 20 seconds ahead and download the tiles only once  based on the predicted FoV, which is absolutely not accurate due to the large prediction interval. Therefore, several frames have almost zero angular resolution and pretty low per-degree quality in this case. In contrast, by progressive downloading, KKT-const predicts FoV and improves the tiles' rates for every frame over $20$ rounds, with the  prediction accuracy becomes more and more accurate over time.

\textbf{Smoothness}: In Fig.~\ref{fig:KKT-const_non-prog_ang_resol_bw5}, \ref{fig:KKT-const_non-prog_frame_quality_per_degree_bw5} and \ref{fig:other_videos_kkt-const_non-prog_frame_quality_per_degree_bw5}, The large quality variation suffered by nonprogressive baseline is also due to the bandwidth variations. At each downloading round of non-progressive baseline, we allocate all the predicted available bandwidth to just one segment consisting of $30$ frames that are about to enter the end of buffer. In contrast, at each downloading round of KKT-const, it updates all the frames in the buffer simultaneously based on the optimization algorithm, which smooths the resulting frame quality evolution dramatically.

% \begin{figure}[htbp]
% \centerline{\includegraphics[width=\linewidth]{figs/KKT-exp_non-prog_ave_frame_quality_bw5.eps}}
% \caption{Frame Quality}
% \label{fig:frame_quality}
% \end{figure}

% render
\begin{figure*}[!t]
    \centering 
    \subfloat[\footnotesize Ground Truth]{\includegraphics[width=.15\linewidth,height=.305\linewidth] {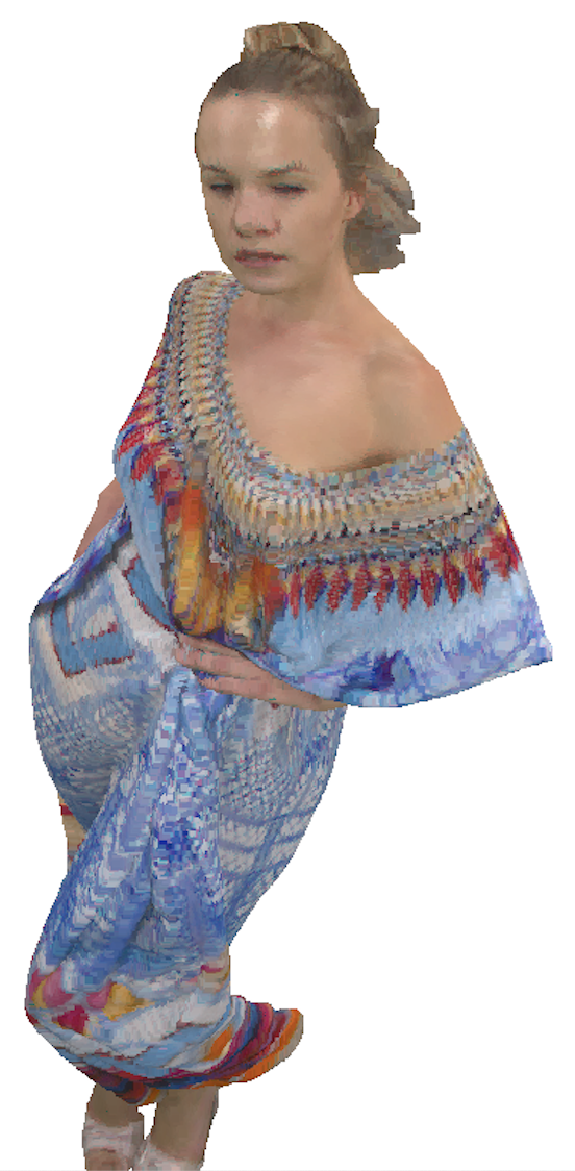}\label{fig:render_gt}
    }
   \hfil
    \subfloat[\footnotesize KKT-exp]{\includegraphics[width=.15\linewidth,height=.305\linewidth] {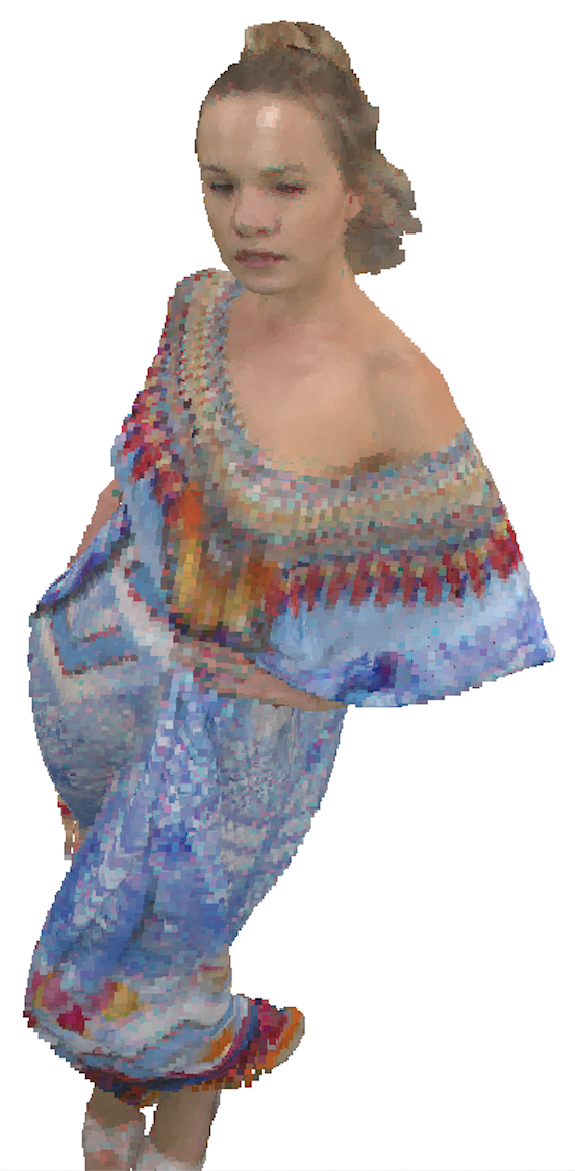}
    \label{fig:render_kktexp}}
    \hfil
    \subfloat[\footnotesize RUMA]{\includegraphics[width=.15\linewidth,height=.305\linewidth]{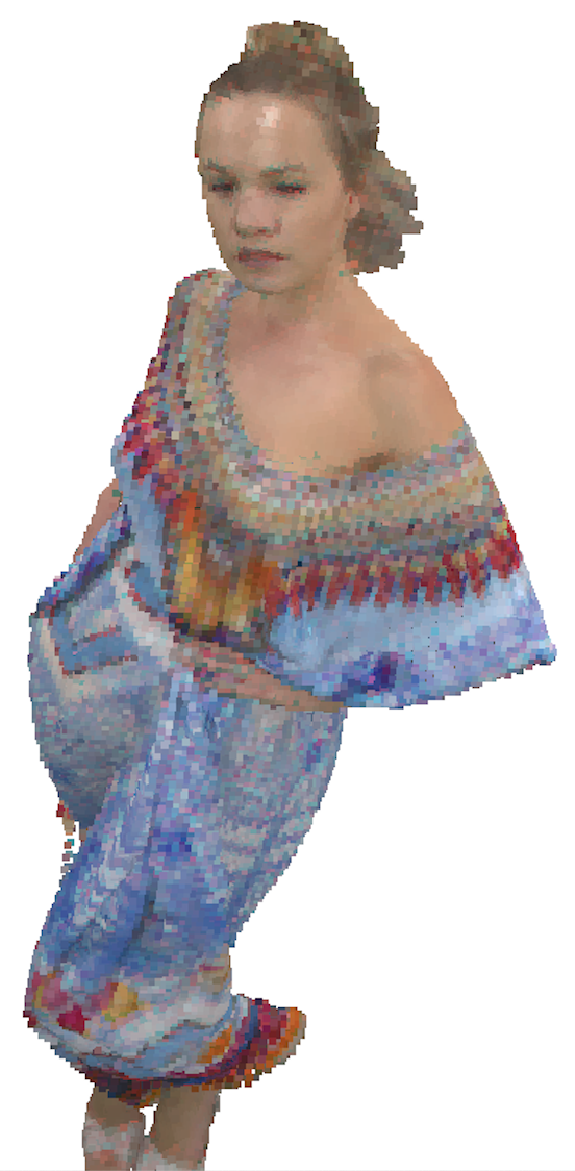}
    \label{fig:render_ruma}}
   \hfil
    \subfloat[\footnotesize Equal Split]{\includegraphics[width=.15\linewidth,height=.305\linewidth] {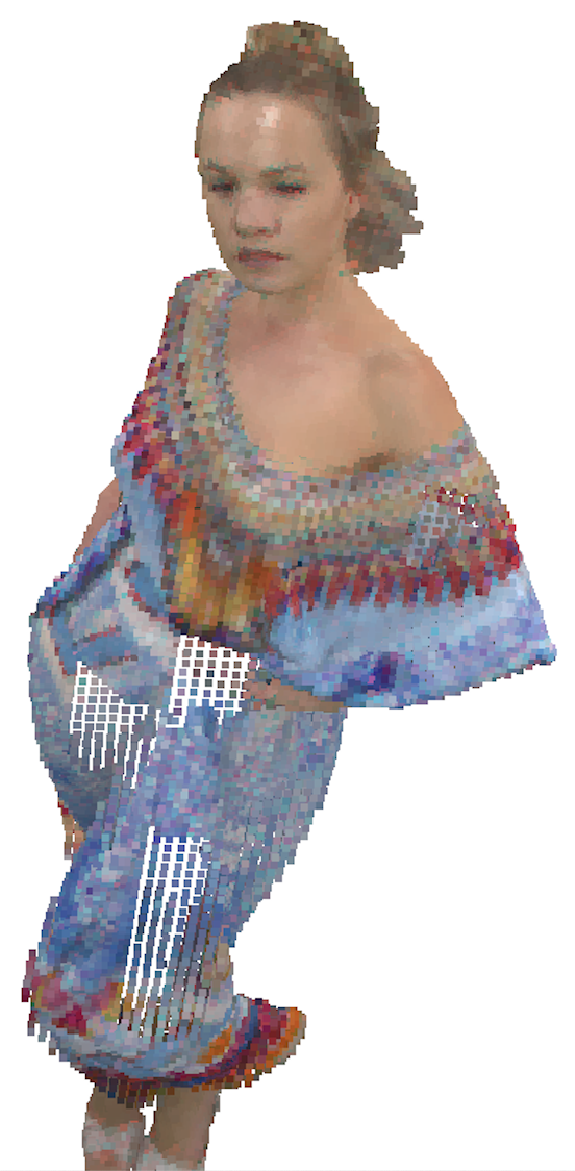}
    \label{fig:render_avr}}
   \hfil
    \subfloat[\footnotesize Non-Progressive]{\includegraphics[width=.15\linewidth,height=.305\linewidth] {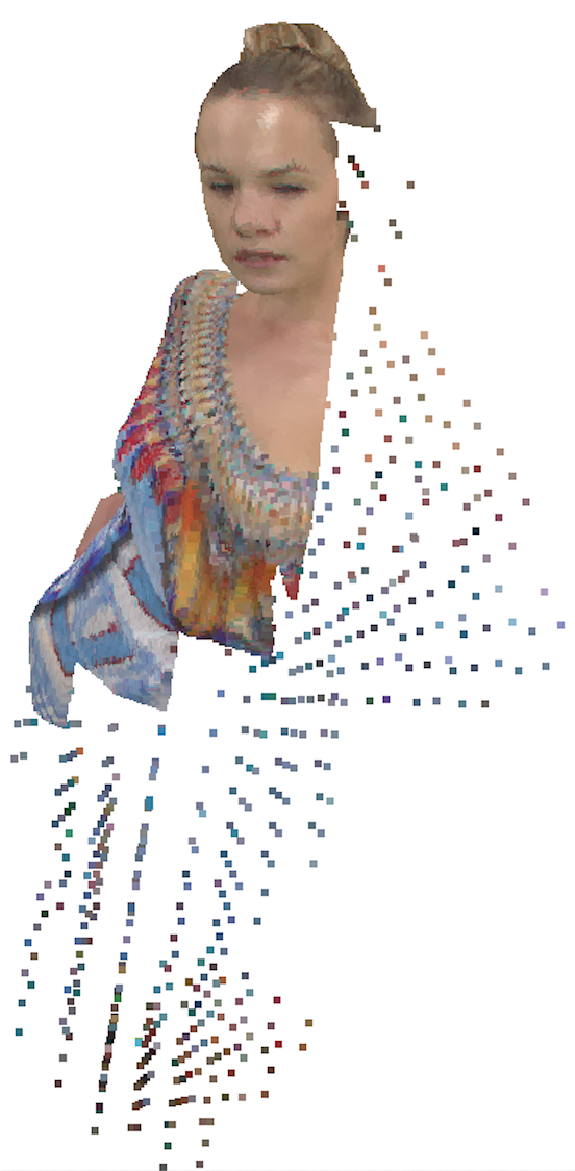}
    \label{fig:render_nonprog}}
    
\caption{Comparison of Rendered Views of Different Baselines.}
\label{fig:render_results}
\end{figure*}

\textbf{Bandwidth Consumption}: At each  downloading time of KKT-const, for the frames farther away from playing, e.g., 20s ahead, since the FoV prediction is not so accurate we just download a base layer for the tiles within the predicted FoV, which doesn't waste too much bandwidth; while for the frames closer to user playback time, we patch the tiles within the more accurately predicted FoV by downloading additional enhancement layers. However, non-progressive baseline allocates all the bandwidth to just one segment based on inaccurate FoV prediction. Therefore, in both Table~\ref{tab:wasted_rates} and Fig.~\ref{fig:KKT-const_non-prog_wasted_rates_bw5} we observe that KKT-const helps save a lot of bandwidth over all the frames. The bandwidth wastage savings would be significant for long PCVs. What is worth mentioning is that in Fig.~\ref{fig:KKT-const_non-prog_wasted_rates_bw5}, we want to fairly compare the wasted rates due to FoV prediction errors between the non-progressive framework and our progressive algorithm. The x-axis extends to $1800$ instead of $2400$, because we remove the comparison results for the first $600$ frames. The FoV prediction for the first $600$ frames are always good as explained in Section~\ref{sec:kkt-const_vs_nonprog}, so the wasted rates of both algorithms are very small for the first $600$ frames. Therefore, we remove them so that the results of the rest frames look more clear.

\begin{figure}[htbp]
\centerline{\includegraphics[width=\linewidth]{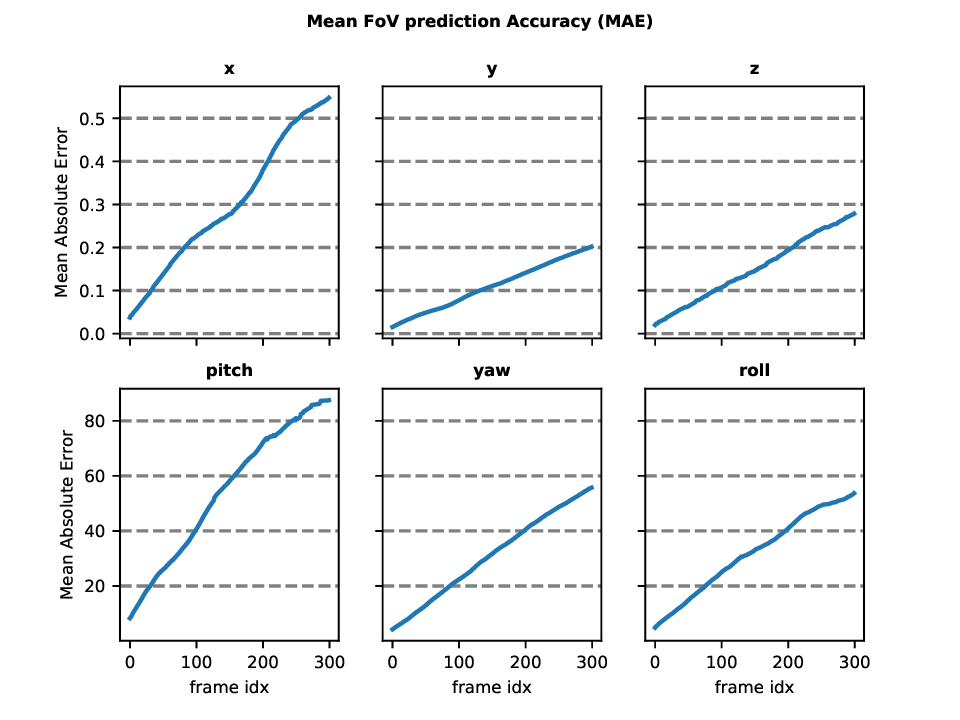}}
\caption{Mean FoV Prediction Accuracy.}
\label{fig:fov_pred_mae}
\end{figure}

\begin{figure*}[!t]
    \centering 
    \subfloat[\footnotesize FoV Predicition Accuracy]{\includegraphics[width=.3\linewidth,height=1.6in] {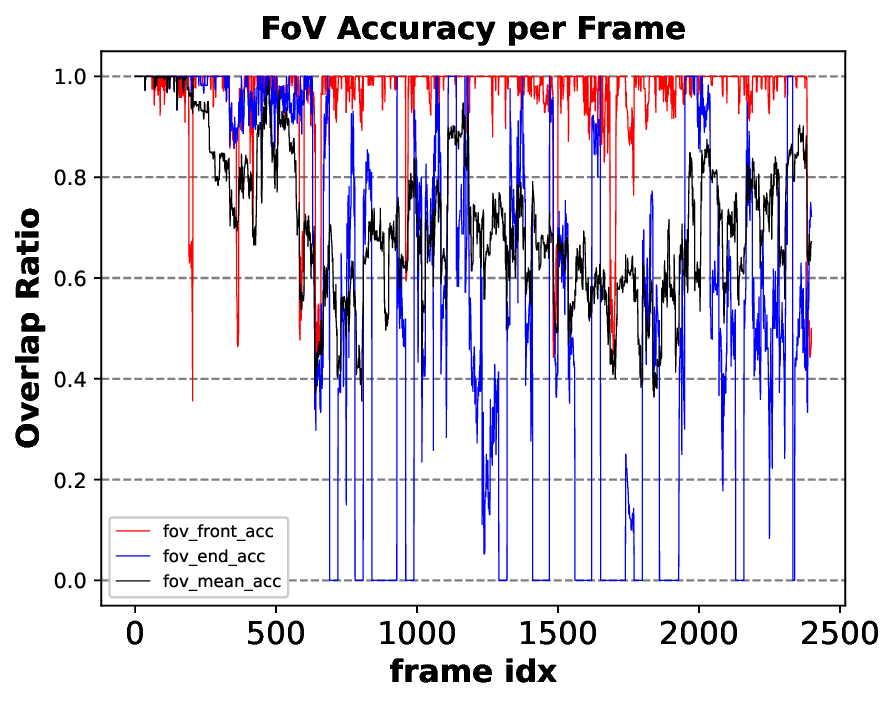}\label{fig:fov_acc_per_fr_overlap_ratio}
    }
    \hfil
    \subfloat[\footnotesize Visual Evidence]{\includegraphics[width=.3\linewidth,height=1.6in] {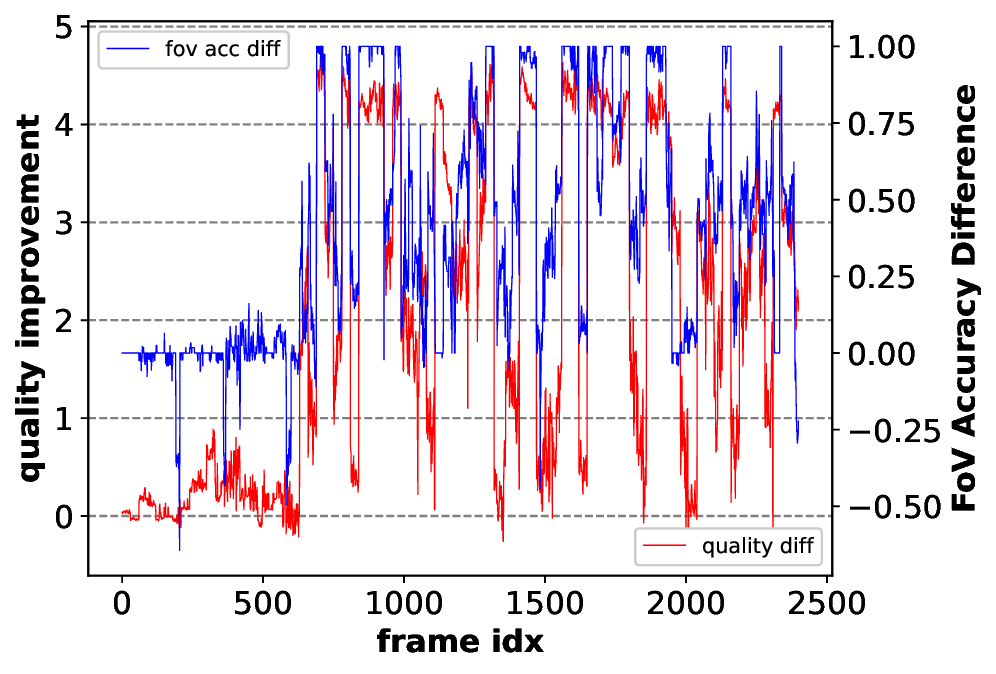}
    \label{fig:KKT-const_non-prog_frame_quality_per_degree_with_fov_acc_bw5}}
    \hfil
    \subfloat[\footnotesize Scatter Plot]{\includegraphics[width=.3\linewidth,height=1.6in]{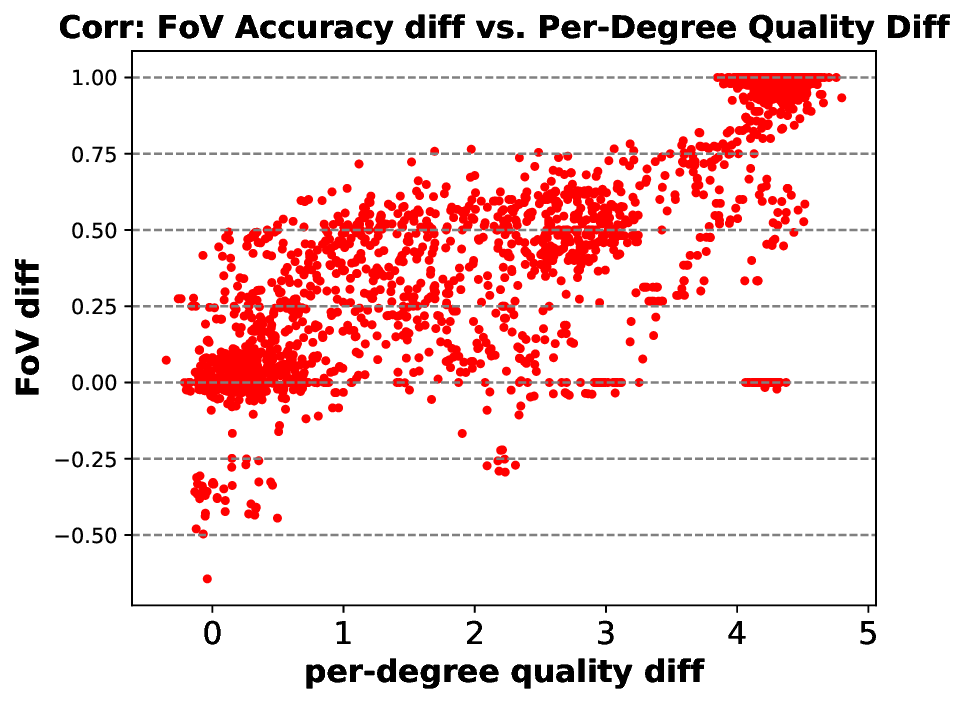}%
\label{fig:corr_fovAcc_diff_KKT-const_non-prog_frame_quality_per_degree_diff_bw5}}
\caption{Correlation between FoV Prediction Accuracy and Quality Improvement: (a) Overlap ratio between predicted and true FoV; (b) VIsual evidence of the correlation; (c) Quality improvement vs. FoV prediction accuracy difference at head and tail of buffer}
\label{fig:FoV vs. Improvement}
\end{figure*}

% psnr & ssim
\begin{table}[htb]%
\centering
    \centering
    % \vspace{-8mm}
    \caption{\small Rendered Quality: Average PSNR and SSIM}
    {
    \begin{tabular}{c|c|c|c|c} \hline
       Metrics   & KKT-exp & RUMA & Equal- & Non-   \\ 
                &  &  & Split & Progressive \\ \hline 
       PSNR & 43.05 & 42.89 & 42.76 & 41.88  \\ \hline
       SSIM & 0.9725 & 0.9695 & 0.9691 & 0.9610 \\  \hline 
       % C &  &  & & \\  \hline 
    \end{tabular}
    }
    \label{tab:psnr_ssim}
   % \vspace{-5mm}
\end{table}

\textbf{Correlation between Quality Improvement and FoV prediction Accuracy}: To verify the motivation of progressive streaming, we zoom in to see the correlation between the quality improvements of progressive streaming and the FoV prediction accuracy. The FoV prediction accuracy in terms of Mean Absolute Error (MAE) with respect to future window sizes (at most $300$ frames ahead) for 6 DoF is shown in Fig.~\ref{fig:fov_pred_mae}. In the progressive downloading process, each frame's FoV is predicted for $20$ times in a 600-frame update window. In Fig.~\ref{fig:fov_acc_per_fr_overlap_ratio}, we show the FoV prediction accuracy in terms of tile overlap ratio with ground-truth user FoV for each frame, where a larger overlap ratio means a more accurate prediction. The blue curve is the first-time prediction when the frame is just download into the buffer and far away from playback time; the red curve is the last-time prediction when the frame is closest to playback time; and black curve is the average accuracy over $20$ predictions. The difference between red and blue curves in Fig.~\ref{fig:fov_acc_per_fr_overlap_ratio} represents the FoV prediction accuracy difference between the front and end position in the buffer.
% We first calculate the FoV prediction accuracy difference between the front position of the buffer (i.e. predicting for frame $t+1$ at frame time $t$ ) and the end of the buffer (i.e. predicting for frame $t+I$, where $I=600$ is update frame window size). The accuracy is represented by the tile overlap ratio between the predicted FoV and the ground-truth FoV: the larger the better. 
We also calculate the quality improvement between KKT-const and non-progressive baseline. When we put the two \textit{difference} curves together in Fig.~\ref{fig:KKT-const_non-prog_frame_quality_per_degree_with_fov_acc_bw5}, we find an obvious correlation between the two: if the FoV prediction at the front of the buffer (closer to user playback time) is \textit{much better} than that at the end of buffer (way ahead of user playback time), the quality improvement of progressive-const against the non-progressive is much larger. The pattern is more obvious in the scatter-plot in Fig.~\ref{fig:corr_fovAcc_diff_KKT-const_non-prog_frame_quality_per_degree_diff_bw5}. The correlation coefficient is $\rho=0.83$.

% \begin{figure}[htbp]
% \centerline{\includegraphics[width=\linewidth]{figs/fov_acc_per_fr.eps}}
% \caption{FoV prediction accuracy in terms of overlap ratio with true user FoV.}
% \label{fig:fov_acc_per_fr_overlap_ratio}
% \end{figure}

% \begin{figure}[htbp]
% \centerline{\includegraphics[width=\linewidth]{figs/kkt-const_non-prog_frame_quality_per_degree_with_fov_acc_bw5.eps}}
% \caption{Clear correlation is observed between the improvement of KKT-const against non-progressive and FoV prediction accuracy difference between the buffer front and the buffer end.}
% \label{fig:KKT-const_non-prog_frame_quality_per_degree_with_fov_acc_bw5}
% \end{figure}

% \begin{figure}[htbp]
% \centerline{\includegraphics[width=\linewidth]{figs/corr_fovAcc_diff_KKT-const_non-prog_frame_quality_per_degree_diff_bw5.eps}}
% \caption{The correlation coefficient is as high as $0.83$.}
% \label{fig:corr_fovAcc_diff_KKT-const_non-prog_frame_quality_per_degree_diff_bw5}
% \end{figure}

\subsubsection{\textbf{KKT-exp V.S. KKT-const}}
\begin{figure}[tbp]
\centering{
\subfloat[\footnotesize Visible Rates]
{\includegraphics[width=0.48\linewidth,height=1.5in]{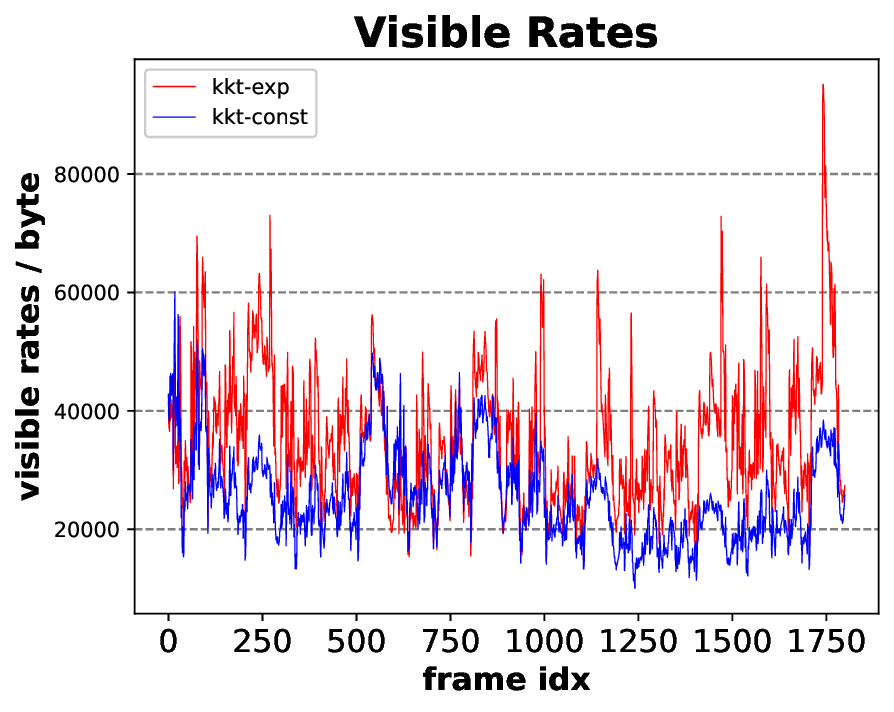}
\label{fig:kkt-exp_kkt-const_visible_rates_bw5}}
\subfloat[\footnotesize Per-Degree Frame Quality]
{\includegraphics[width=0.48\linewidth,height=1.5in]{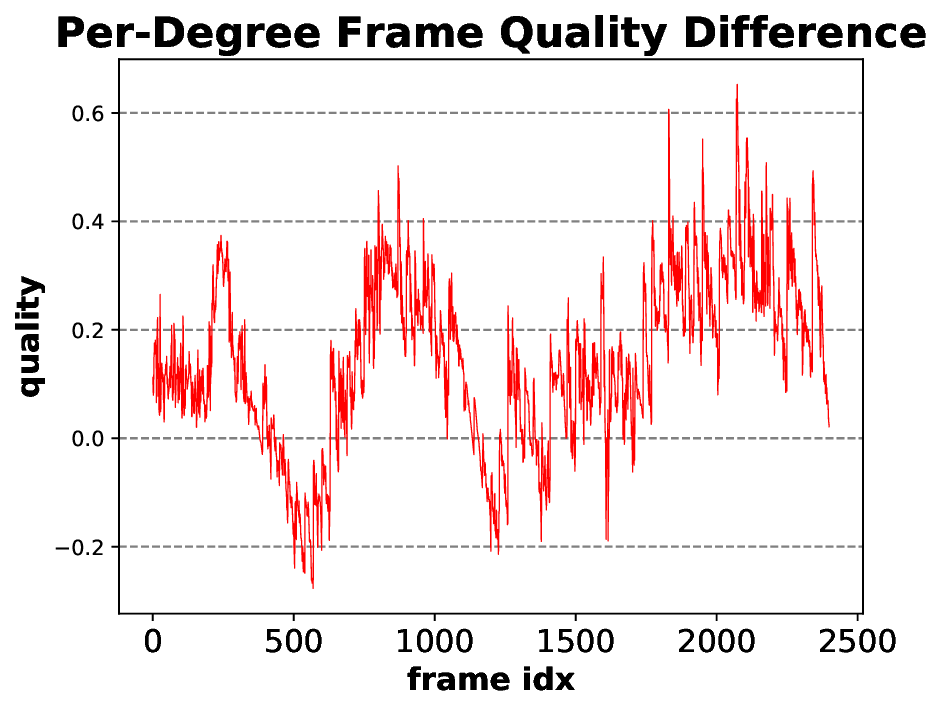}
\label{fig:diff_KKT-exp_KKT-const_frame_quality_per_degree_bw5}
}
\caption{KKT-exp v.s. KKT-const: (a) Visible Rate per Frame; (b) Per-degree Frame Quality Improvement.} \label{fig:kkt-exp v.s. const}}
\end{figure}

We further explore the design space of setting frame weights $\{w_i\}$ in Equation~(\ref{eq:neat_progressive_obj}). The frame weights depend on the confidence about the accuracy of FoV prediction for each frame. Therefore, intuitively, it should decrease as the prediction interval increases. We tried several different weight settings, including linear decreasing, history FoV prediction accuracy based assignment, and exponentially decreasing in terms of prediction interval. We found the exponentially decreasing frame weights performs the best, as shown in Table~\ref{tab:ang_resol}, \ref{tab:frame_quality_per_degree} and \ref{tab:wasted_rates}, as well as in  Fig.~\ref{fig:kkt-exp_kkt-const_visible_rates_bw5} and \ref{fig:diff_KKT-exp_KKT-const_frame_quality_per_degree_bw5} KKT-exp further boosts the per-degree frame quality on top of KKT-const without increasing too much quality variations. From Fig.~\ref{fig:kkt-exp_kkt-const_visible_rates_bw5} and Table~\ref{tab:wasted_rates} we see a large amount of bandwidth is further saved by KKT-exp.

In Fig.~\ref{fig:kkt-exp_kkt-const_visible_rates_bw5}, we can see for some frames the visible rates (i.e. the bits used inside the actual FoV) is more than doubled. That is because the long-term FoV prediction accuracy is low, and downloading tiles outside the actual FoV  wastes a lot of bandwidth. We randomly pick some of these frames as an example to zoom in the improvements of tile quality in different scenarios. We show the improvement distribution over visible tiles in different network conditions, also with different FoV traces that have different distances between the user and the object, since those are two important factors affecting the final viewing quality. 

% KKT-exp vs KKT-const
% \begin{figure}[htbp]
% \centerline{\includegraphics[width=\linewidth]{figs/diff_KKT-exp_KKT-const_frame_quality_per_degree_bw5.eps}}
% \caption{KKT-exp v.s. KKT-const: Per-degree Frame Quality Improvement}
% \label{fig:diff_KKT-exp_KKT-const_frame_quality_per_degree_bw5}
% \end{figure}

% \begin{figure}[htbp]
% \centerline{\includegraphics[width=\linewidth]{figs/kkt-exp_kkt-const_visible_rates_bw5.eps}}
% \caption{KKT-exp v.s. KKT-const: Visible Rate per Frame}
% \label{fig:kkt-exp_kkt-const_visible_rates_bw5}
% \end{figure}

% kkt-exp vs kkt-const: distributions
\begin{figure*}[!t]
    \centering 
    \subfloat[\footnotesize KKT-exp v.s. KKT-const: Tile Rate Improvement Distribution]{\includegraphics[width=.3\linewidth]{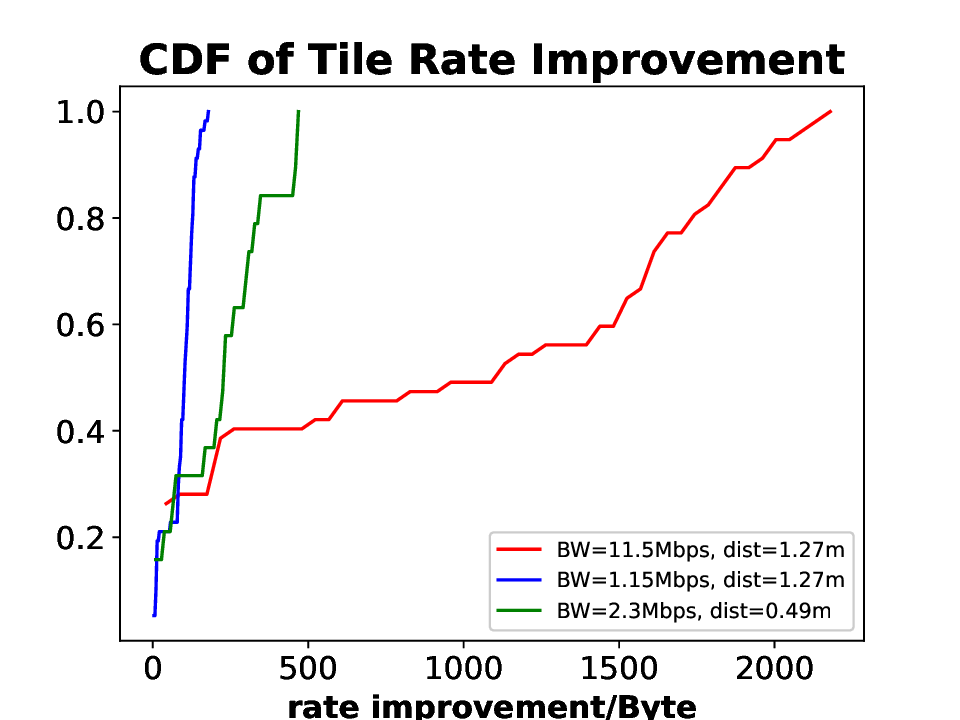}%
    \label{fig:kkt-exp_kkt-const_tileRateDiff_distrFr1470}}
    \hfil
    \subfloat[\footnotesize KKT-exp v.s. KKT-const: Tile Resolution Improvement Distribution]{\includegraphics[width=.3\linewidth]{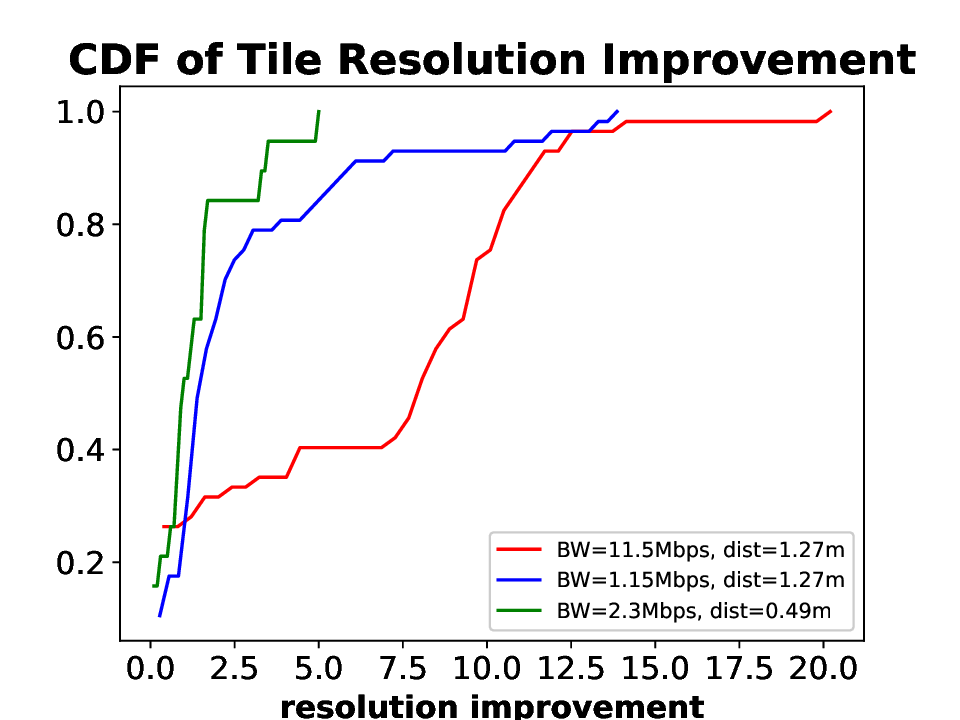}%
    \label{fig:kkt-exp_kkt-const_tileResolDiff_distrFr1470}}
    \hfil
    \subfloat[\footnotesize KKT-exp v.s. KKT-const: Tile Utility Improvement Distribution]{\includegraphics[width=.3\linewidth]{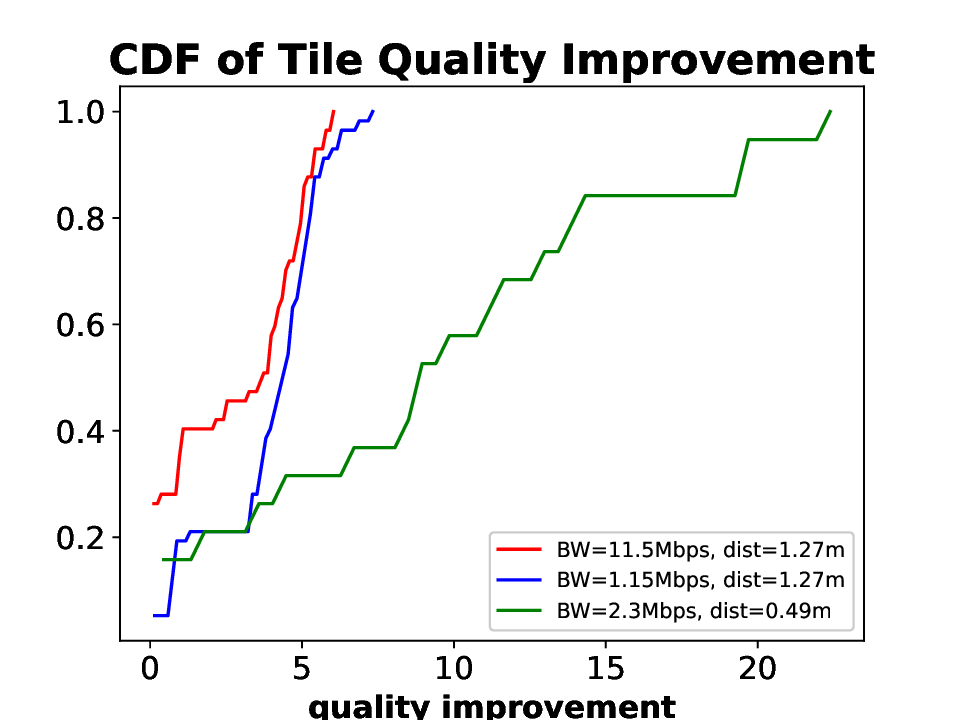}%
    \label{fig:kkt-exp_kkt-const_tileQualityDiff_distrFr1470}}
\caption{KKT-exp vs KKT-const: different network conditions and user-object distances.}
\label{fig:kkt-exp_kkt-const_DistrFr1470}
\end{figure*}

Fig.~\ref{fig:kkt-exp_kkt-const_tileRateDiff_distrFr1470},~\ref{fig:kkt-exp_kkt-const_tileResolDiff_distrFr1470} and~\ref{fig:kkt-exp_kkt-const_tileQualityDiff_distrFr1470} show the CDF of tile rate, resolution and quality improvement respectively under different network conditions and user-object distances. Resolution is the number of points per degree along one dimension in every tile. When bandwidth is high ($11.5Mbps$) and distance is normal ($1.27m$) (red curves), the average visible frame rate of \textit{KKT-const} is $33$ KB, \textit{KKT-exp} is $86$ KB, $2.6x$ times of \textit{KKT-const}. On average, the resolution of \textit{KKT-const} is $13$ pts / degree, while \textit{KKT-exp} is $19$ pts / deg, improved by $46\%$. Also, the average tile quality for \textit{KKT-const} is 28, while \textit{KKT-exp} is 31, improved by $10.7\%$. Although the tiles rate difference is pretty large in this case (red curve) in Fig.~\ref{fig:kkt-exp_kkt-const_tileRateDiff_distrFr1470}, the resolution difference may not be that large because some tiles have higher compression efficiency than others, or they have fewer points.

When bandwidth is small ($1.15Mbps$) and distance is normal ($1.27m$) (blue curves), the average visible frame rate of \textit{KKT-const} is $2979$ Bytes, \textit{KKT-exp} is $8155$ Bytes, $2.7x$ times of \textit{KKT-const}. The average resolution in this case is $3.6$ for \textit{KKT-const} and $6.3$ for \textit{KKT-exp}. Also, the average tile quality for \textit{KKT-const} is 19, while \textit{KKT-exp} is 23, improved by $21.1\%$. Compared to the previous scenario where bandwidth is ten times larger, the tile quality now appears more sensitive to the tile rate. Therefore, although the tile rate improvement ratio is not as large as the previous scenario when bandwidth is high, the tile quality improvement ratio is similar.

We further demonstrate that tile quality is more sensitive to user-object distance change. We keep the bandwidth at a medium level ($2.30Mbps$), while selecting a FoV trace for which the average user-object distance is only $0.49$m. Now the average frame rate improvement is $2.1$ times larger. The resolution improvement is even smaller: $3.6$ for \textit{KKT-const} and $4.9$ for \textit{KKT-exp}. The absolute difference is only less than 2 pts / degree. However, the mean tile quality improvement is the largest among all scenarios as observed from the green curve in Fig.~\ref{fig:kkt-exp_kkt-const_tileQualityDiff_distrFr1470}, for which the absolute improvement of \textit{KKT-exp} compared to \textit{KKT-const} is $9.2$. This result demonstrates that the tile quality is most sensitive to tile rate when user-object distance is small.
Therefore, \textit{KKT-exp} has the greatest improvement compared to \textit{KKT-const} when the user watches the point cloud object very closely.

% quality
% -214.50836845
% -38.943438260000015
% -175.56493018999998
% -11.289914128947368
% -2.0496546452631588
% -9.24025948368421

% rates
% 3889.4182284400003
% 8014.79430635
% -4125.376077909999
% 204.7062225494737
% 421.83127928157893
% -217.12505673210524

% density
% 68.09502357
% 93.37852138999999
% -25.283497819999994
% 3.583948608947368
% 4.914659020526315
% -1.3307104115789472

% \begin{figure}[htbp]
% \centerline{\includegraphics[width=\linewidth]{figs/kkt-exp_kkt-const_tileRateDistrFr1470_bw5.png}}
% \caption{KKT-exp v.s. KKT-const: Tile Rate Distribution}
% \label{fig:kkt-exp_kkt-const_tileRateDistrFr1470_bw5}
% \end{figure}

% \begin{figure}[htbp]
% \centerline{\includegraphics[width=\linewidth]{figs/kkt-exp_kkt-const_tileResolDistrFr1470_bw5.png}}
% \caption{KKT-exp v.s. KKT-const: Tile Resolution Distribution}
% \label{fig:kkt-exp_kkt-const_tileResolDistrFr1470_bw5}
% \end{figure}

% \begin{figure}[htbp]
% \centerline{\includegraphics[width=\linewidth]{figs/kkt-exp_kkt-const_tileQualityDistrFr1470_bw5.png}}
% \caption{KKT-exp v.s. KKT-const: Tile Utility Distribution}
% \label{fig:kkt-exp_kkt-const_tileQualityDistrFr1470_bw5}
% \end{figure}

\subsubsection{\textbf{KKT-exp V.S. Equal Allocation}}
\begin{figure*}[!t]
    \centering 
    \subfloat[\footnotesize Angular Resolution]{\includegraphics[width=.28\linewidth]{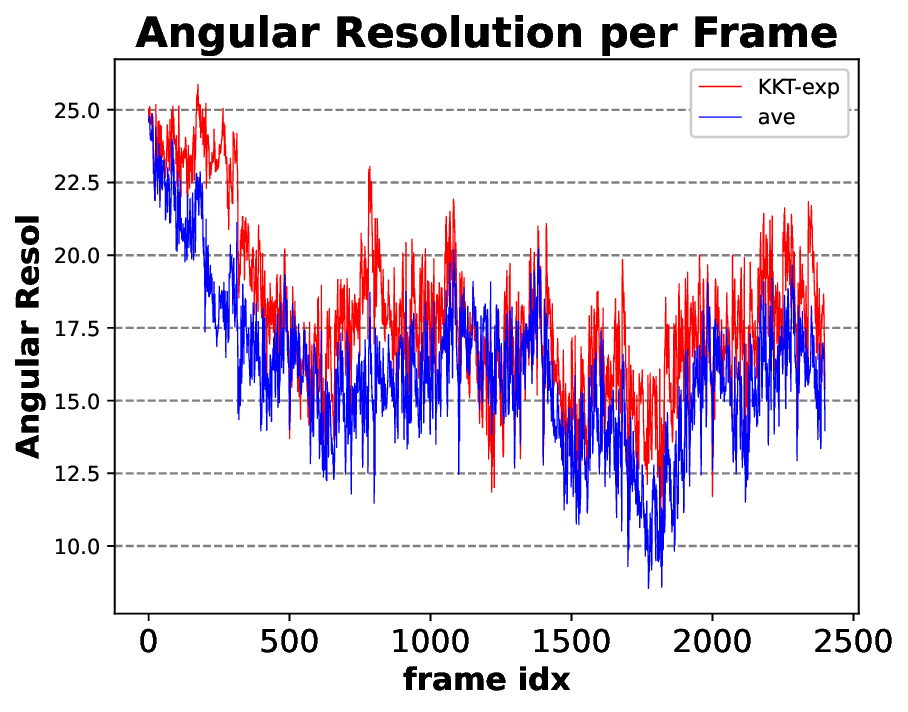}%
    \label{fig:KKT-exp_ave_ang_resol_bw5}}
    \hfil
    \subfloat[\footnotesize Resolution Improvement]{\includegraphics[width=.3\linewidth]{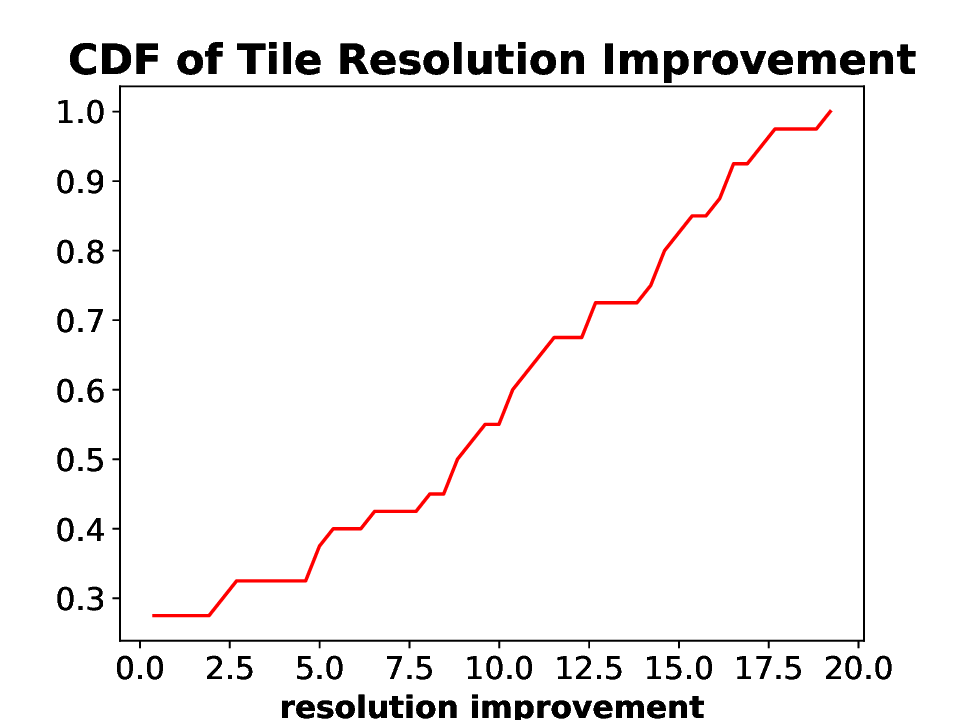}%
    \label{fig:kkt-exp_ave_tileResolDiff_distrFr800_bw5_dist1}}
    \hfil
    \subfloat[\footnotesize Utility Improvement]{\includegraphics[width=.3\linewidth]{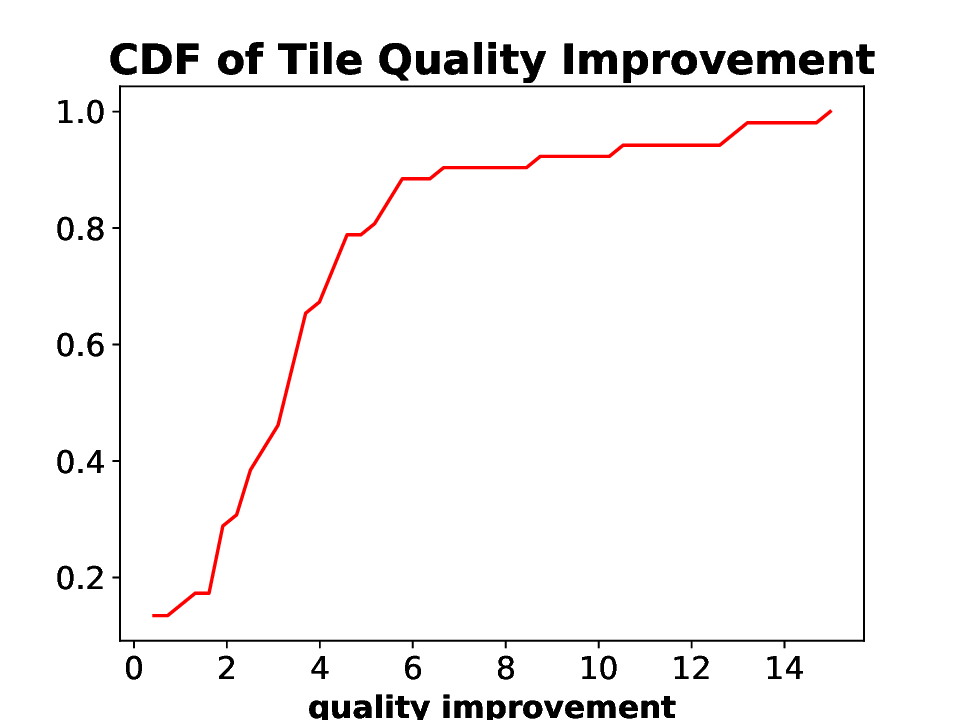}%
    \label{fig:kkt-exp_ave_tileQualityDistrFr2373_bw5_dist1}}
\caption{KKT-exp vs Equal Allocation.}
% \caption{KKT-exp vs Equal Allocation: mean Bandwidth=$\textbf{11.5}$ Mbps, mean distance=$\textbf{1.27}$m.}
\label{fig:kkt-exp_ave_Distr_bw5_dist1}
\end{figure*}

% \textbf{???add a distribution comparison???}
To demonstrate the superiority of KKT-based  tile rate allocation for PCV streaming, we design an equal allocation baseline, where the predicted available bandwidth is equally split over all the tiles within the predicted FoV of every frame. We show the angular resolution per frame comparison to the equal allocation baseline in Fig. \ref{fig:KKT-exp_ave_ang_resol_bw5}, and resolution improvement CDF distribution in Fig~\ref{fig:kkt-exp_ave_tileResolDiff_distrFr800_bw5_dist1}, where KKT-exp reaches $25pts/^\circ$ against $17pts/^\circ$ of Equal allocation. Additionally, we show mean tile quality improvement CDF in Fig.~\ref{fig:kkt-exp_ave_tileQualityDistrFr2373_bw5_dist1}.

% \begin{figure}[htbp]
% \centerline{\includegraphics[width=\linewidth]{figs/KKT-exp_ave_ang_resol_bw5.eps}}
% \caption{KKT-exp v.s. Equal Allocation: Angular Resolution per Frame}
% \label{fig:KKT-exp_ave_ang_resol_bw5}
% \end{figure}

% kkt-exp vs equal allocation: distributions

\subsubsection{\textbf{KKT-exp V.S. RUMA}}
\begin{figure}[tbp]
\centering{
\subfloat[\footnotesize Per-Degree Frame Quality\label{fig:kkt-exp_ruma_frame_quality_per_degree_bw5}]
{\includegraphics[width=0.46\linewidth]{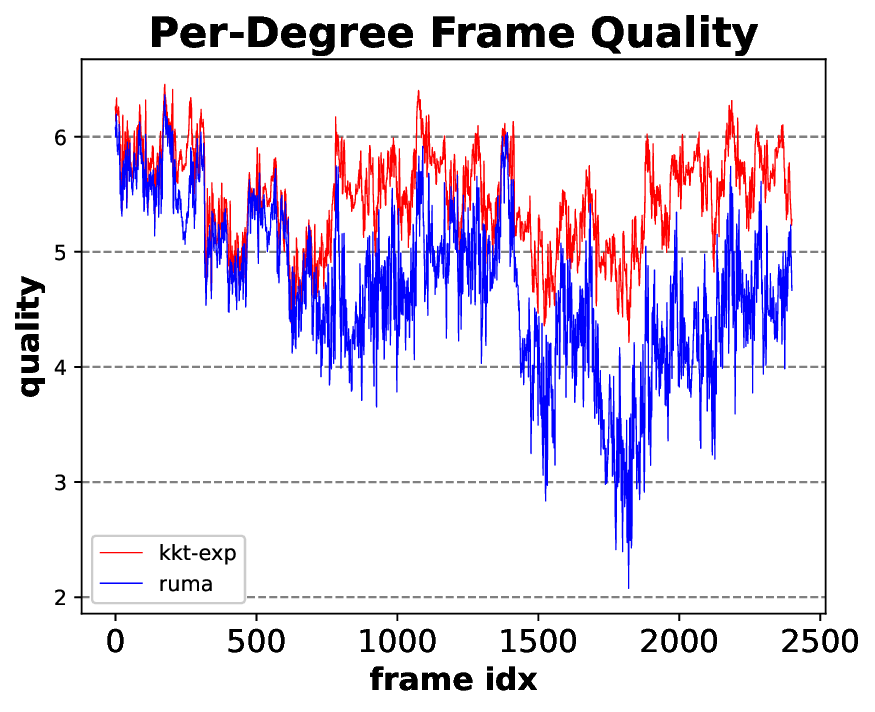}}
\subfloat[\footnotesize Tile Utility Distribution\label{fig:kkt-exp_ruma_tileQualityDistrFr1470_bw5_dist1_lod}]
{\includegraphics[width=0.52\linewidth]{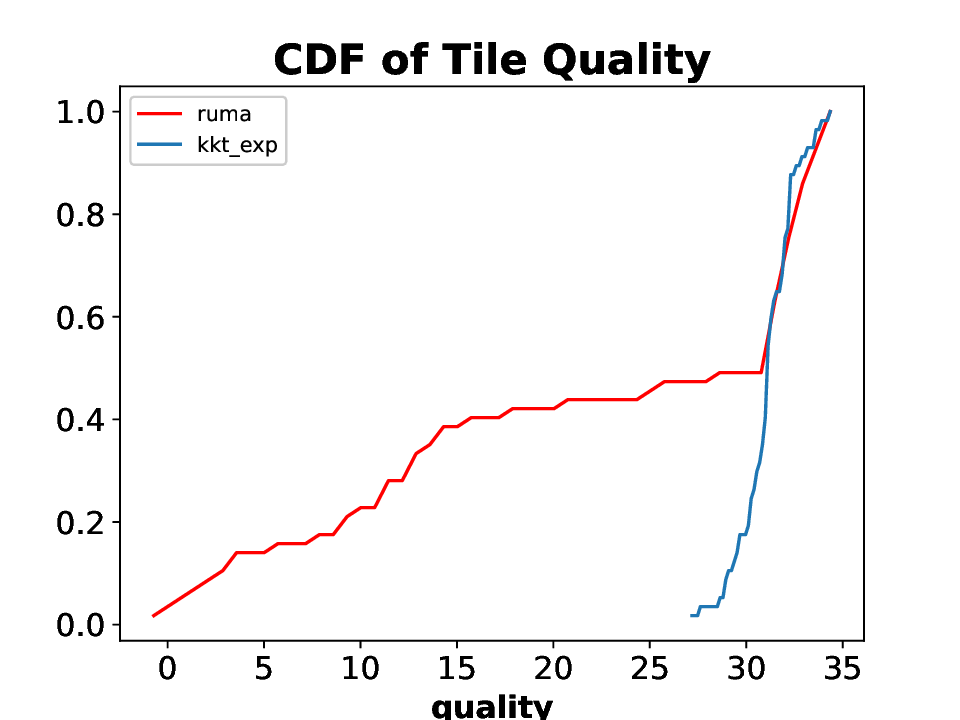}}
\caption{KKT-exp v.s. RUMA} \label{fig:kkt_exp_vs_ruma}}
\end{figure}

We now compare KKT-exp with the state-of-the-art rate allocation algorithm RUMA~\cite{park2019rate}. In Fig.~\ref{fig:kkt-exp_ruma_frame_quality_per_degree_bw5}, we see the per-degree frame quality evolution over time. The mean per-degree quality of KKT-exp is $5.42$ against $4.46$ of RUMA, an improvement by $22\%$ on average.

% \begin{figure}[htbp]
% \centerline{\includegraphics[width=\linewidth]{figs/kkt-exp_ruma_frame_quality_per_degree_bw5.eps}}
% \caption{KKT-exp v.s. RUMA: Frame Quality per Degree}
% \label{fig:kkt-exp_ruma_frame_quality_per_degree_bw5}
% \end{figure}

% \begin{figure}[htbp]
% \centerline{\includegraphics[width=\linewidth]{figs/kkt-exp_ruma_tileQualityDistrFr1470_bw5_dist1_lod.png}}
% \caption{KKT-exp v.s. RUMA: Tile Utility Distribution}
% \label{fig:kkt-exp_ruma_tileQualityDistrFr1470_bw5_dist1_lod}
% \end{figure}

In Fig.~\ref{fig:kkt-exp_ruma_tileQualityDistrFr1470_bw5_dist1_lod}, we zoom in some frames whose quality improvement is larger than the others. We can see that due to the greediness of RUMA, even if some tiles have achieved the highest quality, it fails to optimize the rate allocation that should consider all the tiles of interests. Also, the intermediate steps of RUMA estimate the marginal tile utility by finite difference, which is not accurate enough to ensure an optimal solution. Instead, we first use KKT to solve the continuous rate allocation problem, which guarantees the optimality because we use the true marginal utility in the calculation. We then round the solution to each tile's discrete rate version.

\subsubsection{Visual Results}
\label{sec:render_result}
In Fig.~\ref{fig:render_results}, we show the example reandered views of different baselines. The artifacts of ``Equal Split" and ``Non-Progressive'' can be easily observed. Non-progressive result looks weird because it downloads each frame only once when the the frame is about to be download into buffer and the FoV prediction is very inaccurate. For state-of-art ``RUMA'', when we zoom in we can see the obvious artifacts around hair, right face and chest as well as left thigh, while there's no obvious artifacts for ``KKT-exp''. We also show the average PSNR and SSIM results in Table~\ref{tab:psnr_ssim}. We conclude that the proposed KKT-Condition based progressive downloading strategy performs better than the baselines in terms of visual results, which proves the rationality of the proposed quality model.

\section{Acknowledgements}
The project was partially supported by USA National Science Foundation under award number CNS-2312839.

\section{Conclusion}
\label{sec:conclusion}
As volumetric video is on the rise, we explore the design space of point cloud video streaming which is challenging due to its high communication and computation requirements. By relying on the inherent scalability of octree-based point cloud coding, we proposed a novel {\it sliding-window based progressive streaming framework} to gradually refine the spatial resolution of each tile as its playback time approaches and FoV prediction accuracy improves. In this way, we managed to balance needs of long streaming buffer for absorbing bandwidth variations and short streaming buffer for accurate FoV prediction. We developed an analytically optimal algorithm based on the Karush–Kuhn–Tucker (KKT) conditions to solve the tile rate allocation problem.
% \item We further formulate the rate allocation problem as an optimal control decision making problem and solve it with iterative Linear Quadratic Regulator (iLQR), which further enhances the Quality of Experience (QoE) performance compared to KKT-based algorithm due to its calculation of future impacts of every action.
We also proposed a novel tile rate-utility model that explicitly considers the viewing distance to better reflect the true user QoE. In the future, we will not only consider maximizing the total tile utility, but also controlling the quality variations between consecutive frames, which leads to a more complicated non-concave objective function, and introduces stronger coupling between rate allocations on adjacent frames. We will apply nonlinear optimal control techniques, e.g.  iterative Linear Quadratic Regulator (iLQR), to solve the optimal rate allocation problem.

\bibliographystyle{IEEEtran}
\bibliography{ref}

\end{document}